\documentclass[twocolumn]{article}
\usepackage[utf8]{inputenc}
\usepackage[english]{babel}
\usepackage[T1]{fontenc}
\usepackage{amsmath}
\usepackage{graphicx}
\usepackage{booktabs}
\usepackage{xurl}
\usepackage{xstring}

\newcommand{\keywords}[1]{\textbf{Key words:} #1}

\begin{document}

\title{Neutral Atom Quantum Computing: Principles, Routes, Progress, and Challenges}
\author{
  Junchao Wang$^{1}$\thanks{Corresponding author.}, 
  Zeyuan Wang$^{1}$, Lei Li$^{1}$, Feng Wang$^{1}$, Shibo Liang$^{1}$, and Keduo Yan$^{1}$\\
  $^{1}$Information Engineering University
}
\maketitle

\begin{abstract}
Neutral atom quantum computing utilizes laser-trapped neutral atoms as qubits and realizes quantum logic gate operations through Rydberg-state interactions. In recent years, it has become one of the most vibrant directions in quantum computing hardware. This paper systematically reviews the working principles of neutral-atom quantum computers, including qubit encoding, atom trapping and manipulation, Rydberg states and interactions, the Rydberg blockade quantum gate mechanism, and atom rearrangement with reconfigurable architectures. The mainstream technical routes are surveyed, represented by optical tweezer arrays combined with Rydberg interactions, optical lattice schemes, and dipole trap arrays. A panoramic review is provided of domestic and international research progress from theoretical foundations in 2000 to the latest achievements in 2026, including thousand-qubit-scale systems, logical qubits, and quantum error correction experiments. Key breakthroughs are highlighted, such as the 6100-atom qubit array, continuous operation of a 3000-qubit system, quantum simulation of the Kitaev honeycomb model, toric code error correction demonstrations, encoding rates exceeding 1/2, and fault-tolerant architectures. The core bottlenecks are analyzed in depth, including the scalability--fidelity trade-off, engineering implementation of quantum error correction, atom loss and mid-circuit replenishment, laser system industrialization, control electronics scalability, and long-distance quantum interconnection. This paper aims to provide a systematic reference for academic research and technological development in this field.
\end{abstract}


\section{Introduction}

The physical realization of quantum computing is one of the most challenging frontier topics in contemporary physics. Over the past two decades, researchers have explored a variety of physical systems as carriers of quantum bits, including superconducting Josephson junctions, trapped ions, quantum dots, topological superconducting Majorana fermions, photons, and neutral atoms. Each platform has its own advantages and disadvantages: superconducting qubits feature fast gate operations but require cryogenic environments and complex chip fabrication processes; trapped-ion qubits offer extremely long coherence times and high gate fidelities but are limited in scalability by ion-chain crosstalk and heating; neutral-atom quantum computing, by contrast, has significant advantages in scalability, long coherence times, all-to-all connectivity, and the absence of dilution refrigerators, and has developed particularly rapidly in the past five years.

The core idea of neutral-atom quantum computing is to use laser fields (optical tweezers or optical lattices) to trap neutral atoms, encode the internal atomic states as qubits, and excite atoms to highly excited Rydberg states, leveraging the extremely strong dipole--dipole or van der Waals interactions between Rydberg atoms to implement quantum logic gates. Compared with superconducting and trapped-ion platforms, the neutral-atom platform has several key advantages. First, atoms are identical natural quantum systems; no fabrication process is needed to ensure qubit uniformity. Second, optical tweezer arrays operate in room-temperature vacuum chambers without dilution refrigerators, reducing cryogenic engineering complexity. Third, atoms can be moved dynamically during computation by optical tweezers, enabling dynamic reconfiguration and, in principle, arbitrary topological connectivity. Fourth, three-dimensional spatial arrangement is a natural physical route to scaling, and experiments have already demonstrated parallel manipulation of thousands of atoms. Since 2020, neutral-atom quantum computing has entered a period of explosive development---from Atom Computing's 1180-qubit array in 2023 to the first logical quantum processor in 2024, and further to an ~11000-atom array in 2026 and continuous operation of 3000 qubits. These breakthroughs indicate that the field is advancing from fundamental research toward practical fault-tolerant quantum computing.

Limitations of existing reviews: In 2010, Saffman et al. published a review in Reviews of Modern Physics that systematically elaborated the physical principles and early experimental progress of the field, becoming a foundational reference. However, in the more than ten years since that review, neutral-atom quantum computing has experienced leapfrog development from proof of principle to thousand-qubit-scale logical quantum processors. In particular, breakthrough progress has been made in quantum error correction, fault-tolerant architectures, and industrialization, which have not yet been systematically covered by Chinese-language reviews.

The contributions of this paper are as follows: (1) it covers the latest landmark achievements from 2021--2026, including logical quantum processors, encoding-rate breakthroughs, and fault-tolerant architectures; (2) it sorts out the domestic and international industrial landscape and commercialization progress; (3) it adds quantitative comparisons of technical routes and quantitative analyses of bottlenecks. The organization of this paper is as follows: Section 2 explains the working principles of neutral-atom quantum computing; Section 3 reviews mainstream technical routes and compares them quantitatively; Section 4 chronicles domestic and international research progress; Section 5 summarizes industrialization progress at home and abroad; Section 6 analyzes the core bottlenecks currently faced; and Section 7 provides an outlook and concluding remarks.

\section{Working Principles}
The physical basis of neutral-atom quantum computing involves multiple aspects, including qubit encoding, atom trapping and manipulation, Rydberg-state interactions, and Rydberg-blockade-based quantum gates. This section begins with qubit encoding schemes and then introduces atom trapping and manipulation technologies, the physical properties of Rydberg states and interaction mechanisms, the Rydberg blockade quantum gate principle, and atom rearrangement with reconfigurable architectures.

\subsection{Qubit Encoding}

In neutral-atom quantum computing, qubit information is stored in the internal energy levels of atoms (see Fig. 1). The most commonly used encoding scheme employs the hyperfine states of alkali-metal atoms (such as rubidium-87 and cesium-133) in the ground state. Taking rubidium-87 as an example, its 5S1/2 level splits into two hyperfine sublevels, F = 1 and F = 2, under nuclear-spin coupling, with an energy spacing of about 6.8 GHz, which can serve as stable |0⟩ and |1⟩ states. The coherence times of these hyperfine states can reach several seconds or even longer, far exceeding typical quantum gate operation times (on the microsecond scale), providing conditions for executing complex quantum algorithms.

In addition to hyperfine-state encoding, nuclear-spin-state encoding of alkaline-earth-metal atoms (such as strontium-87 and ytterbium-171) has attracted attention in recent years. Alkaline-earth-metal atoms have a 1S0 ground state and a 3P0 metastable state. Because the total electronic angular momentum of the 1S0 and 3P0 states is zero, the coupling between nuclear spin and the electronic environment is extremely weak, enabling coherence times of nuclear-spin states to reach tens of seconds or even minutes. This "clock-state" encoding scheme has been adopted by companies such as Atom Computing.

Furthermore, alkaline-earth-metal atoms without nuclear spin, such as strontium-88, can use transitions between the 1S0 and 3P2 (or 3P0) states for encoding; their metastable-state lifetimes are as long as hundreds of seconds, supporting high-fidelity quantum operations. The choice of different atomic species involves trade-offs among coherence time, laser cooling efficiency, and Rydberg-state characteristics.
\begin{figure}[htbp]
\centering
\includegraphics[width=0.45\textwidth]{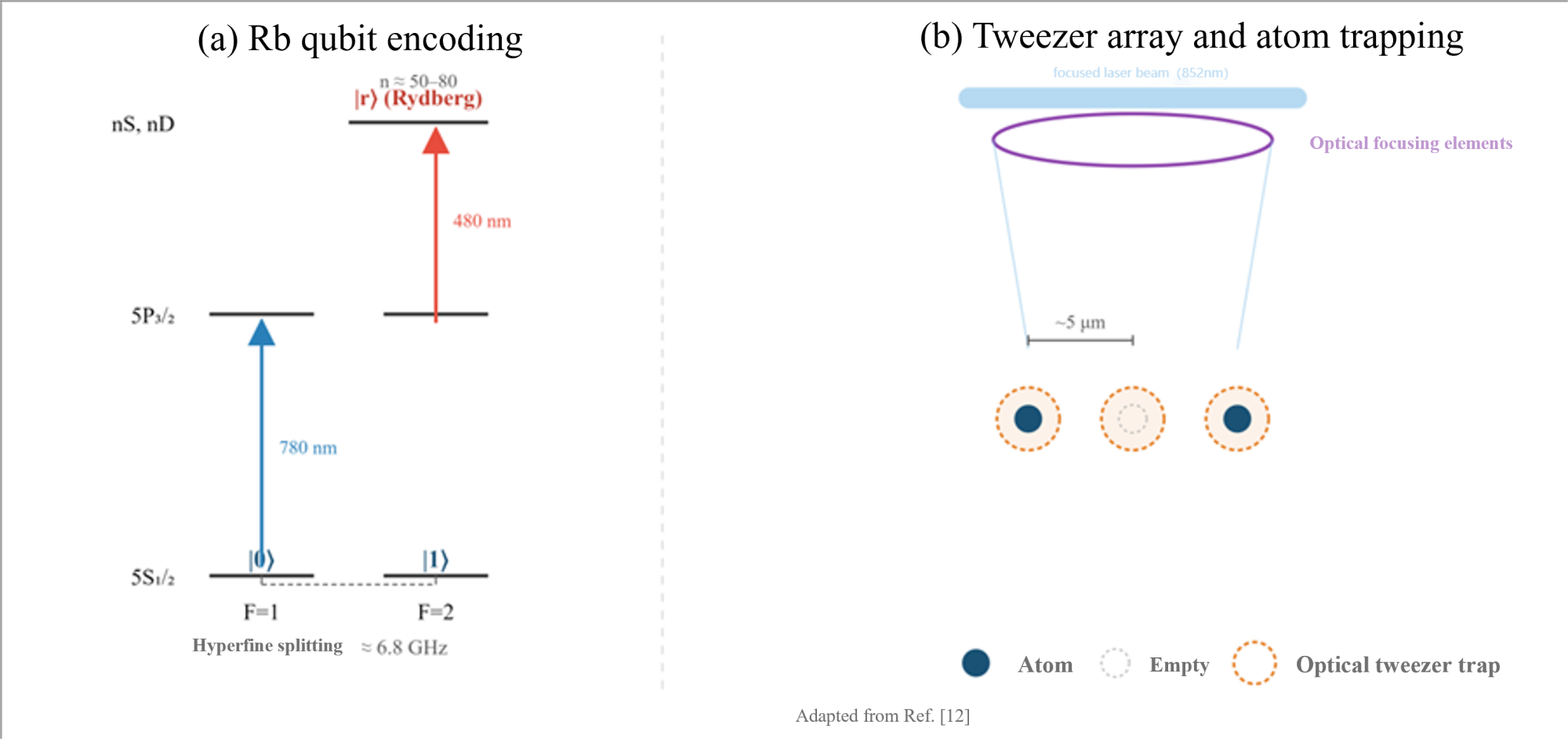}
\caption{Qubit encoding and optical tweezer array of neutral atoms. (a) Energy-level structure of a rubidium-87 atom. The hyperfine states $F=1$ and $F=2$ of the $5S_{1/2}$ ground state encode $|0\rangle$ and $|1\rangle$, respectively, and a two-photon transition (780 nm + 480 nm) excites the atom to a high principal-quantum-number $nS$ or $nD$ Rydberg state $|r\rangle$. (b) Schematic of an optical tweezer array. A focused laser beam is split by optical focusing elements into a periodic trap array; each trap confines one atom, and dashed circles indicate vacancies (adapted from Ref. [12].)}
\label{fig:1}
\end{figure}

\subsection{Atom Trapping and Manipulation}
The trapping of neutral atoms relies on the dipole force arising from light–atom interactions. When the laser frequency is lower than the atomic transition frequency (red-detuned), atoms are attracted to the region of maximum light intensity; when the laser frequency is higher than the transition frequency (blue-detuned), atoms are repelled to the region of minimum light intensity. Optical tweezers formed by highly focused laser beams can trap individual atoms in tiny regions.

Optical tweezer arrays are currently the mainstream trapping scheme for neutral-atom quantum computing. A single laser beam is split into hundreds to thousands of independent foci by acousto-optic deflectors (AODs) or spatial light modulators (SLMs), with each focus trapping one atom to form one-dimensional, two-dimensional, or three-dimensional programmable atomic arrays. The typical spacing between tweezers is several micrometers, which ensures independence of atoms in the ground state (ground-state interactions are negligible) while allowing sufficiently strong interactions after Rydberg excitation. A key advantage of optical tweezer arrays is their reconfigurability: the position of each tweezer can be independently controlled, so the geometric configuration of the atomic array can be dynamically changed during computation, enabling arbitrary topological connectivity.

Optical lattices use the interference of multiple laser beams to form periodic potential well arrays and can trap large numbers of atoms in three-dimensional space. The advantages of optical lattices lie in their natural high density and uniformity: each lattice site can trap one atom, and in principle tens of thousands or more qubits can be arranged. However, independent addressing of individual sites in an optical lattice is relatively difficult, and the lattice configuration is hard to reconfigure dynamically. The French company Pasqal adopts the optical lattice scheme and carries out quantum simulation and quantum computing through an analog–digital hybrid mode.

Dipole trap arrays use micro-fabricated optical elements (such as microlens arrays) to generate dense trap arrays and represent an intermediate scheme between optical tweezers and optical lattices. Microlens arrays can produce highly uniform trap arrays, but their independent addressing capability is weaker than that of AOD-driven optical tweezer schemes.

Regardless of the trapping scheme, atoms must first be laser-cooled (usually by a magneto-optical trap, MOT) to microkelvin temperatures before being transferred to optical tweezers or optical lattices. A typical experimental workflow is: MOT capture of cold atomic cloud → random loading of single atoms into tweezers → imaging to detect atom occupancy → atom rearrangement to eliminate vacancies → execution of quantum operations → readout of results.

\subsection{Rydberg States and Interactions}
Rydberg states are highly excited states in which the valence electron of an atom is excited to a large principal quantum number $n$. In a Rydberg state, the atomic orbital radius is proportional to $n^2$ and can reach hundreds of nanometers or even micrometers, far larger than that of a ground-state atom. This huge atomic size gives Rydberg atoms a series of extreme physical properties: extremely large polarizability (proportional to $n^7$), extremely long radiative lifetimes (proportional to $n^3$, reaching hundreds of microseconds to milliseconds), and extremely strong interatomic interactions.

There are two main forms of interaction between two Rydberg atoms. When the two atoms are in the same Rydberg state, the interaction is dominated by van der Waals forces, with potential energy proportional to $1/R^6$ (where $R$ is the interatomic distance) and strength reaching 10--100 MHz (at $R \sim 10 \, \mu m$). When dipole--dipole interactions are brought into resonance by an external electric field or by selecting specific Rydberg states, the interaction potential is proportional to $1/R^3$ and is stronger. Both interactions far exceed typical laser--atom coupling strengths (several MHz), sufficient to implement fast quantum logic gates.

Rydberg atoms are usually excited through two-photon transitions: an ultraviolet or blue laser excites the atom from the ground state to an intermediate level, and an infrared or red laser couples the intermediate level to the target Rydberg state. The two-photon scheme can effectively avoid the heating effects of the high-energy ultraviolet light required for single-photon transitions and can achieve high-efficiency, low-decoherence Rydberg excitation by choosing appropriate intermediate levels and laser parameters.

\subsection{Rydberg Blockade and Quantum Gates}

Rydberg blockade is the core physical mechanism of neutral-atom quantum computing (see Fig. 2). Its basic principle is as follows: when one atom is excited to a Rydberg state, the strong interaction between Rydberg atoms causes the Rydberg-state energy level of nearby atoms to shift significantly (by an amount much larger than the linewidth of the excitation laser), making these atoms unable to be resonantly excited. This "blockade" effect means that within a certain range (the blockade radius $R_b$), at most one atom can be excited to the Rydberg state.

Based on Rydberg blockade, two-qubit entangling gates can be realized. Taking the CZ gate (controlled-phase gate) as an example, the typical implementation procedure is:

1. Apply a $\pi$ pulse to the control qubit to attempt to excite it to the Rydberg state. If the control qubit is in the $|1\rangle$ state, it is successfully excited; if it is in the $|0\rangle$ state (not encoded as a Rydberg-coupled state), it is unaffected.

2. Apply a $2\pi$ pulse to the target qubit to attempt to excite it to the Rydberg state. If the control qubit has already been excited (is in the Rydberg state), the target qubit cannot be excited due to the blockade effect, and its quantum state acquires a phase factor; if the control qubit has not been excited, the target qubit normally undergoes a $2\pi$ Rabi oscillation and returns to its original state without acquiring an extra phase.

3. Apply a $\pi$ pulse again to the control qubit to de-excite it from the Rydberg state back to the ground state.

The final result is that only when both the control and target qubits are in the $|1\rangle$ state does the target qubit acquire a $\pi$ phase, that is, the controlled-phase operation $|11\rangle \rightarrow -|11\rangle$ is realized, equivalent to a CZ gate. Combined with single-qubit rotations, CNOT gates and universal quantum gate sets can be further constructed.

The fidelity of Rydberg blockade quantum gates is affected by many factors, including spontaneous emission from Rydberg states, laser phase noise, interaction fluctuations caused by atomic thermal motion, and leakage errors due to incomplete blockade. Through optimal control techniques to optimize pulse shapes, using atomic dark states to reduce scattering, and improving Rydberg excitation and atomic cooling schemes, two-qubit gate fidelity has increased from about 58\% (Bell-state fidelity) in 2010 to more than 99.5\% in 2023, exceeding the theoretical threshold of surface-code quantum error correction (about 1\%). However, it should be noted that the threshold theorem requires all operations in the error-correction circuit (including measurement and feedback) to satisfy the threshold conditions. Actual error-correction circuits must also consider measurement errors and atom loss, so there is still a gap toward practical fault-tolerant computation.

\begin{figure}[htbp]
\centering
\includegraphics[width=0.45\textwidth]{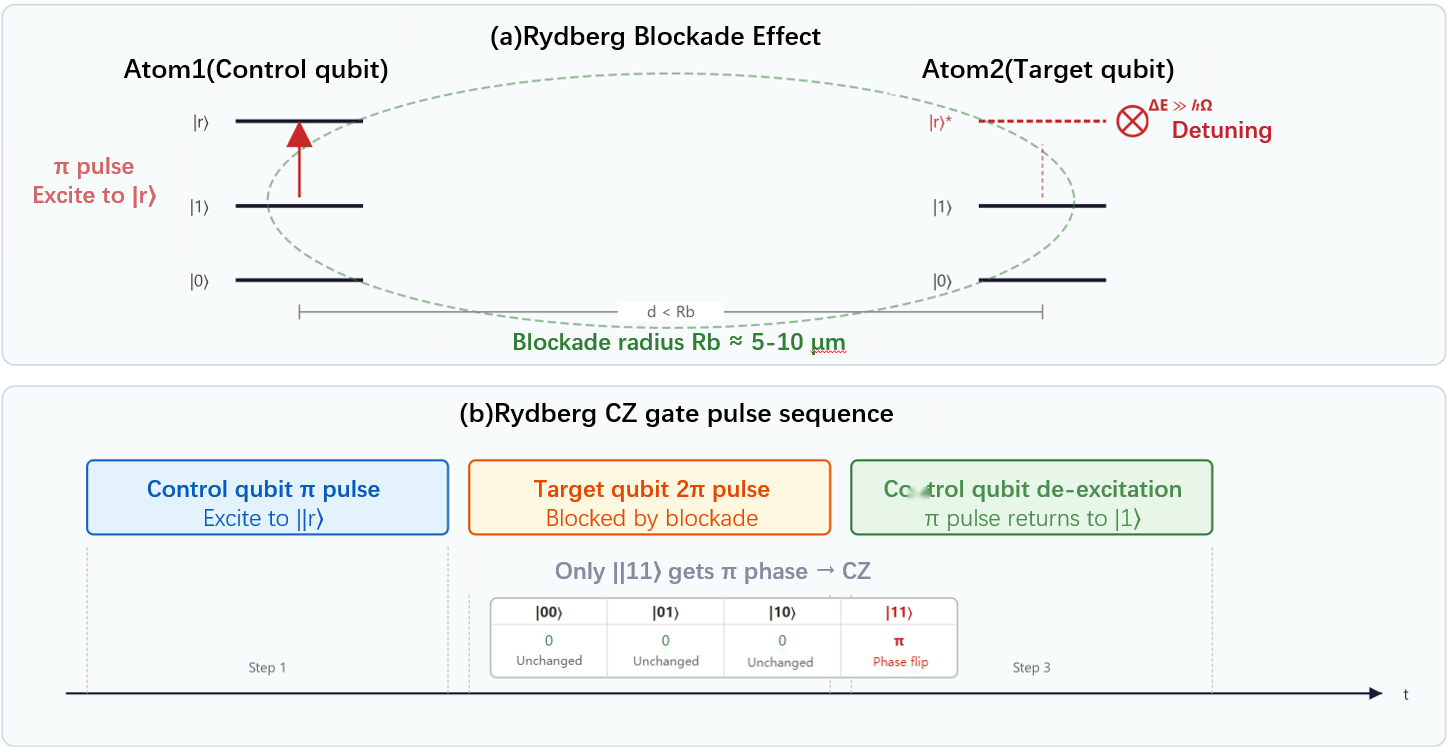}
\caption{Rydberg blockade quantum gate mechanism. (a) Blockade effect: after atom 1 is excited to the Rydberg state $|r\rangle$, the $|r\rangle$ energy level of atom 2 within the blockade radius $R_b$ shifts significantly by $\Delta E \gg \hbar\Omega$ and cannot be resonantly excited (red dashed line + $\times$). (b) CZ gate pulse sequence: a $\pi$ pulse excites the control qubit, a $2\pi$ pulse on the target qubit accumulates phase, and the control qubit is de-excited; only the $|11\rangle$ state acquires a $\pi$ phase (adapted from Refs. [14,15].)}
\label{fig:2}
\end{figure}

\subsection{Atom Rearrangement and Reconfigurable Architecture}

A typical experimental workflow of neutral-atom quantum computing is shown in Fig. 3. A unique advantage is that atoms can be dynamically rearranged during computation by moving optical tweezers. In the initial loading stage, each trap in the optical tweezer array is randomly occupied by one atom with a probability of about 50\%, resulting in many vacancies. After identifying the vacancy positions by imaging detection, mobile optical tweezers are used to transfer atoms from excess positions to vacant positions, allowing defect-free atomic arrays to be constructed within millisecond timescales.

Atom rearrangement is not only used to initialize defect-free arrays but, more importantly, enables reconfigurable qubit connectivity topologies. During quantum algorithm execution, two qubits that need to interact can be brought within the blockade radius by moving optical tweezers, the entangling gate is executed, and then they are separated again. This ``move--entangle--separate'' operation mode allows arbitrary pairs of qubits to establish connections, realizing all-to-all interconnection without fixed nearest-neighbor coupling. This is an important architectural advantage of the neutral-atom platform over superconducting and semiconductor qubits.

In 2025, Pan Jianwei's team at the University of Science and Technology of China used artificial intelligence to drive a high-speed spatial light modulator and implemented an array-scale-independent constant-time rearrangement scheme, successfully constructing defect-free two-dimensional and three-dimensional arrays of 2024 atoms within 60 milliseconds, setting a new world record. The key innovation of this scheme is to move all atoms simultaneously rather than one by one, so that the rearrangement time does not increase with array scale, making future arrays of tens of thousands of atoms possible.

\begin{figure}[htbp]
\centering
\includegraphics[width=0.45\textwidth]{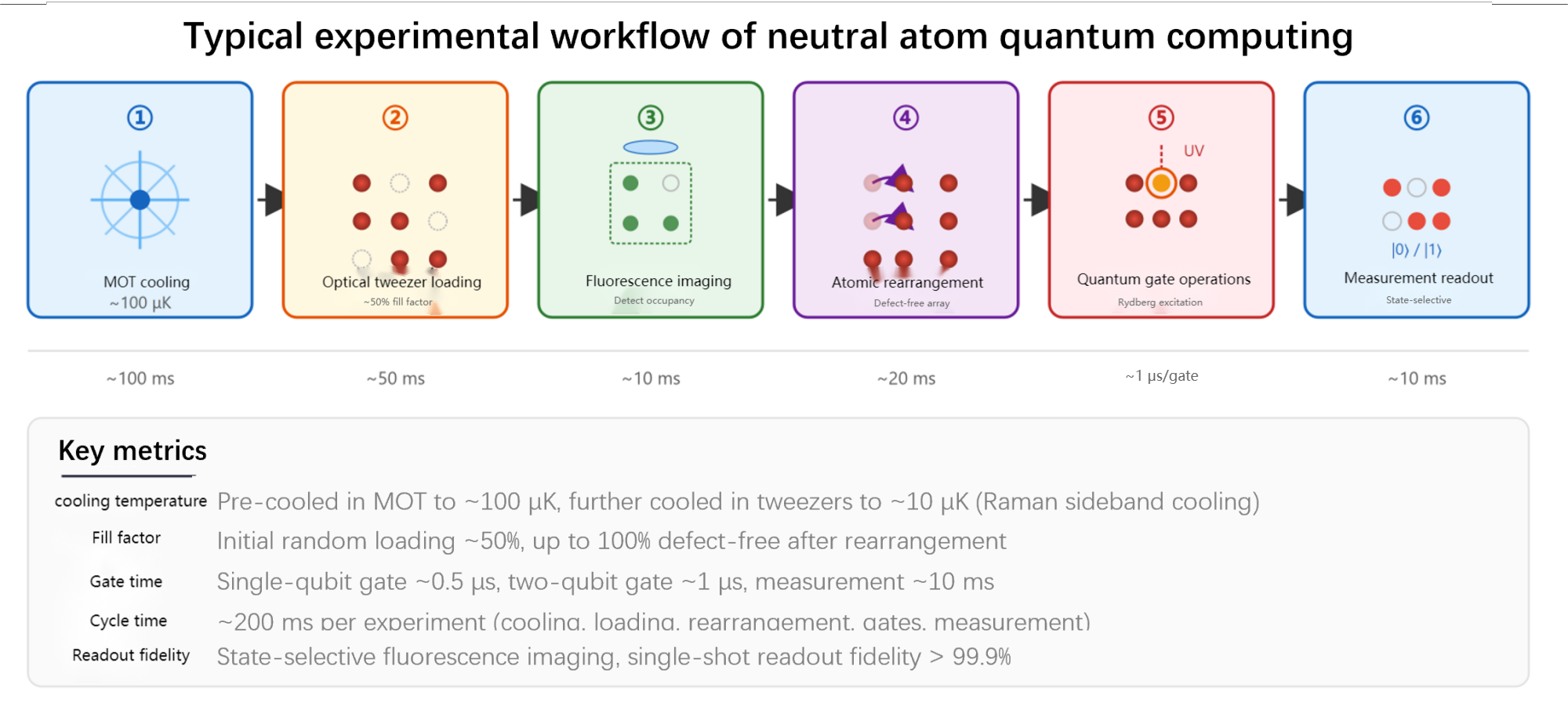}
\caption{Typical experimental workflow of neutral atom quantum computing. The process starts with magneto-optical trap (MOT) capture of cold atoms, followed by random loading of single atoms into optical tweezers ($\sim 50\%$ filling), fluorescence imaging detection, atom rearrangement to eliminate vacancies, Rydberg-laser quantum gate operations, and finally fluorescence readout of the quantum state. A typical cycle takes $\sim 100$ ms, with key fidelity parameters noted (adapted from Refs. [1,12,17].)}
\label{fig:3}
\end{figure}

\section{Mainstream Technical Routes}

At present, neutral-atom quantum computing has formed three technical routes: optical tweezer arrays combined with Rydberg interactions as the mainstream, supplemented by optical lattices and dipole trap arrays. These schemes differ in scaling capability, gate fidelity, reconfigurability, and engineering maturity. This section introduces the principles and representative progress of the three routes one by one, presents a quantitative comparison, and discusses key enabling technologies such as Rydberg-state preparation, non-destructive readout, quantum error correction codes, and atom replenishment.

\subsection{Optical Tweezer Arrays + Rydberg Interactions}

The combination of optical tweezer arrays and Rydberg interactions is the current mainstream technical route for neutral-atom quantum computing and is also the scheme adopted by leading companies such as QuEra Computing (USA) and Atom Computing (USA). The core advantage of this route lies in combining the reconfigurability of optical tweezers with the universal quantum gate capability of Rydberg blockade, enabling gate-model quantum computing. As shown in Fig. 4, a spatial light modulator (SLM) or acousto-optic deflector (AOD) splits a single laser beam into hundreds to thousands of focused spots, forming independently addressable two-dimensional optical tweezer arrays. Each tweezer traps one neutral atom (usually rubidium-87), and the qubit is encoded on the hyperfine ground state of the atom. When executing quantum gates, target atoms are pumped to Rydberg states by two-photon excitation, and entangling gates between adjacent atoms within the blockade radius are realized by the Rydberg blockade effect. The movability of optical tweezers allows atoms to be dynamically rearranged in the array, enabling reconfigurable qubit connectivity topologies, which is the core advantage of this scheme over the optical lattice scheme.

QuEra Computing was incubated from the Lukin group at Harvard University and adopts the rubidium-87 atom and optical tweezer array scheme. In October 2023, QuEra, together with Harvard University and the Massachusetts Institute of Technology, achieved a two-qubit entangling gate fidelity of 99.5\% on a 60-atom neutral-atom quantum computer; this fidelity exceeds the theoretical threshold of surface-code quantum error correction (about 1\%, i.e., 99\% fidelity; the specific threshold varies between 0.57\% and 1.1\% depending on the noise model). In January 2024, Bluvstein et al. reported in \textit{Nature} the first programmable logical quantum processor. The system used 280 rubidium atoms to build a three-zone processor with storage, entanglement, and readout zones, and executed fault-tolerant quantum algorithms on 48 logical qubits through transversal gate operations, with logical error rates up to ten times lower than physical qubit error rates. In September 2025, QuEra, together with Harvard University and Yale University, published the ``Algorithmic Fault Tolerance'' (AFT) framework in \textit{Nature}. By combining transversal operations and correlated decoding, it reduced runtime overhead by about a factor of $d$ (more than 30 times in simulations) while maintaining exponential decay of logical error rates, and can shorten the execution time of large-scale logical algorithms by 10--100 times when mapped onto reconfigurable atom arrays.

Atom Computing adopts the alkaline-earth-metal strontium atom and optical tweezer array scheme, using nuclear-spin-state encoding to achieve extremely long coherence times. In October 2023, Atom Computing announced that its second-generation system achieved more than 1180 physical qubits, setting a record for the number of qubits on a neutral-atom platform at that time. In November 2024, Atom Computing, in cooperation with Microsoft, reported in \textit{Nature} the creation and entanglement of 24 logical qubits using the Bacon--Shor code, which was the largest-scale logical qubit entanglement at that time. In the same experiment, 28 logical qubits ran the Bernstein--Vazirani algorithm with higher accuracy than unencoded physical qubits, demonstrating for the first time the advantage of active quantum error correction on a commercial neutral-atom platform. Atom Computing's roadmap plans to deliver the Magne system in 2027, using 1225 physical qubits to generate 50 logical qubits.

\begin{figure}[htbp]
\centering
\includegraphics[width=0.45\textwidth]{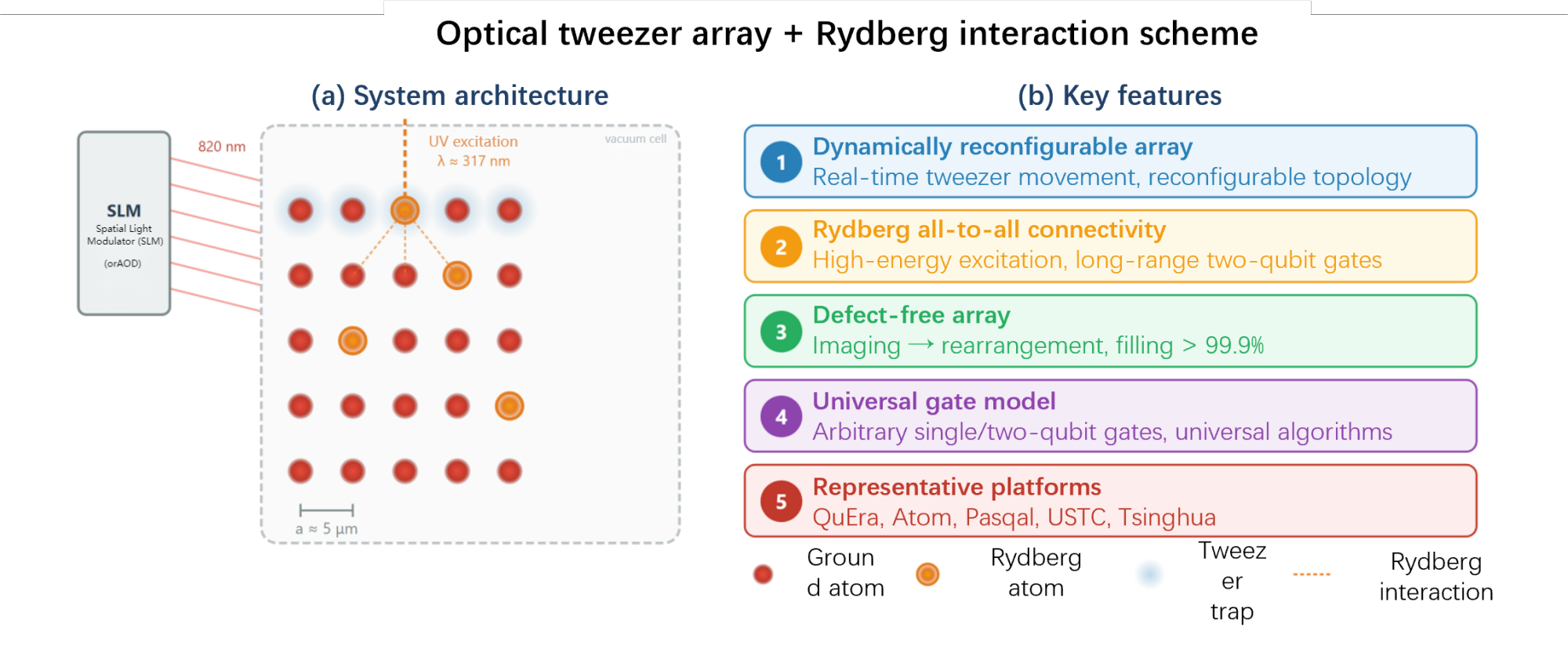}
\caption{Schematic of optical tweezer array with Rydberg interaction. A single laser beam is split by a spatial light modulator (SLM) or acousto-optic deflector (AOD) into hundreds to thousands of focused spots, forming an independently addressable two-dimensional optical tweezer array (adapted from Refs. [2,3].)}
\label{fig:4}
\end{figure}

\subsection{Optical Lattice Scheme}

The optical lattice scheme is represented by the French company Pasqal. Pasqal was incubated by the Browaeys and Lahaye teams at the Institut d'Optique in France and adopts the rubidium atom and optical lattice (or optical tweezer array hybrid) scheme, emphasizing the analog--digital hybrid quantum computing mode. As shown in Fig. 5, an optical lattice is formed by the interference of two counterpropagating laser beams, creating a standing-wave field with periodic potential wells at the nodes; atoms are trapped at the minima of the potential wells to form large-scale uniform arrays. Unlike the optical tweezer scheme, the trap spacing of an optical lattice is determined by the laser wavelength and has extremely high uniformity, making it suitable for parallel quantum simulation. In analog mode, all atoms are simultaneously excited to Rydberg states, and the evolution of quantum many-body systems is directly simulated using Ising-type Hamiltonians, making it suitable for solving combinatorial optimization and quantum simulation problems. In digital mode, universal quantum algorithms are executed through serialized single-qubit and two-qubit gates.

In June 2024, Pasqal announced that it had achieved a quantum processor with more than 1000 atoms in a single loading. Between 2024 and 2025, Pasqal deployed its 100+ qubit Orion Beta system at the French GENCI high-performance computing center and the German Jülich Research Centre, becoming one of the first quantum computers integrated and deployed on-site in supercomputing facilities. Pasqal's roadmap plans to achieve 10,000 qubits by 2028 and 100 logical qubits by 2029.

The advantage of the optical lattice scheme lies in its ability to leverage existing cold-atom experimental technology to achieve large-scale atomic arrangements, and the uniformity of the lattice is beneficial for parallel quantum simulation. Its challenges are that independent addressing and dynamic reconfiguration capabilities are weaker than those of the optical tweezer scheme, and need to be addressed by optical-tweezer-assisted or new optical engineering technologies.

\begin{figure}[htbp]
\centering
\includegraphics[width=0.45\textwidth]{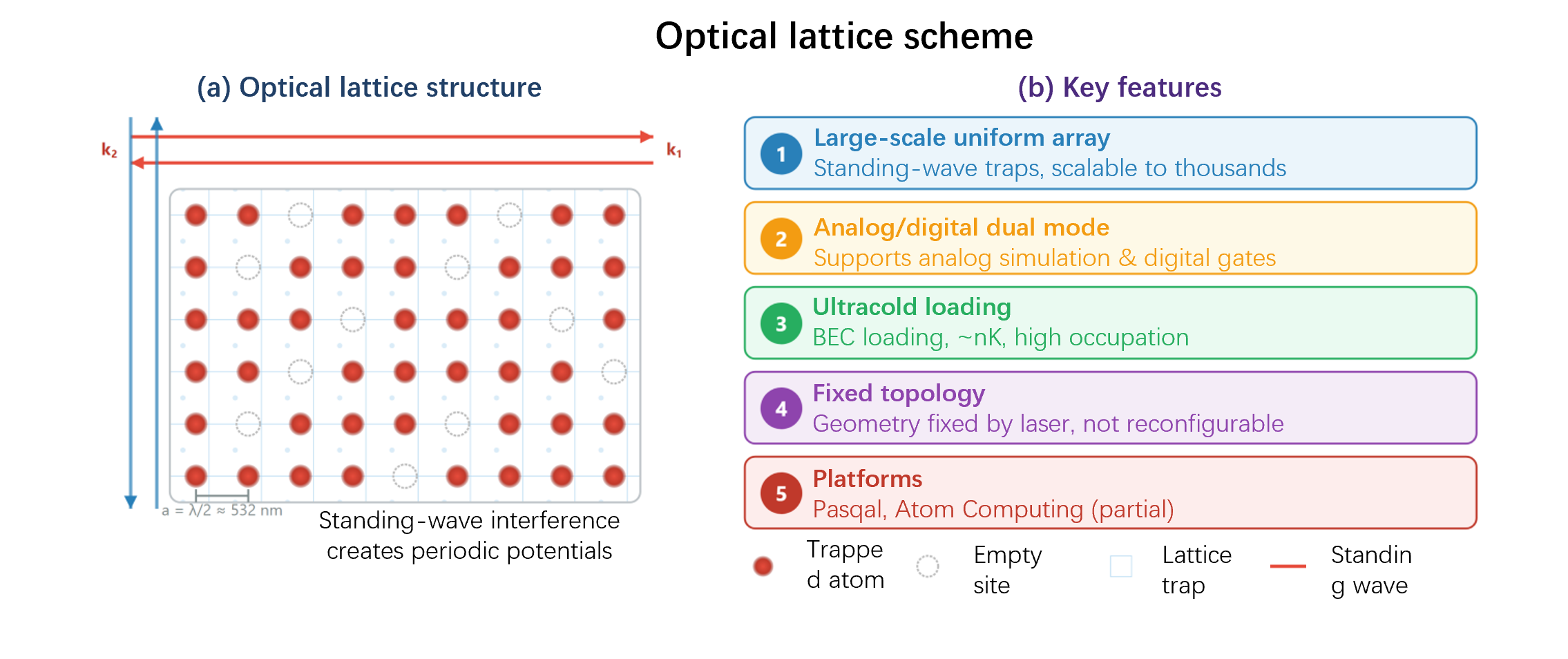}
\caption{Schematic of optical lattice scheme and computation modes. Two counterpropagating laser beams form a standing wave, creating periodic potential wells at the nodes where atoms are trapped to form a large-scale uniform array (adapted from Ref. [6].)}
\label{fig:5}
\end{figure}

\subsection{Dipole Trap Arrays and Alternative Schemes}

Dipole trap arrays use micro-fabricated optical elements such as microlens arrays to generate dense trap arrays and represent a technical route between optical tweezers and optical lattices. As shown in Fig. 6a, the microlens array scheme uses a micro-fabricated lens array to focus a broad laser beam into a dense array of dipole traps; its advantage lies in the high uniformity of trap spacing, but independent addressing requires additional optical systems. The integrated photonic chip scheme (Fig. 6b) uses on-chip optical waveguides and grating couplers to emit light vertically upward, forming trapping potential wells above the chip, representing the frontier direction for miniaturization and engineering of neutral-atom quantum computers. In addition, the fiber array scheme processes the fiber end face into a microlens, which can also achieve compact dipole trap arrays. In the early exploration of dipole trap arrays, the Saffman group proposed a blue-detuned dipole trap scheme, using blue-detuned laser light to trap atoms at the weakest light intensity to reduce decoherence caused by the light field, and implemented Rydberg-blockade controlled-NOT gates and Bell-state entanglement in a two-dimensional cesium atom array, laying an experimental foundation for the subsequent development of large-scale dipole trap arrays. These schemes overlap with optical tweezer schemes in trapping and manipulation technologies and supplement the diversity of neutral-atom quantum computing platforms.

In recent years, the development of integrated photonics has made it possible to manufacture compact dipole trap arrays on chips, promoting the miniaturization and engineering of neutral-atom quantum computers.

In addition, some alternative schemes are under exploration. For example, the alkaline-earth-metal atom scheme using clock-state encoding (such as Atom Computing's strontium platform) obtains extremely long coherence times through nuclear spin states ($T_2$ echo reaching 40 seconds) while implementing quantum gates through coupling between metastable states and Rydberg states. Another example is the scheme using color centers in two-dimensional materials or solid-state atomic defects as qubits; although not belonging to the traditional neutral-atom category, it intersects with neutral-atom schemes in trapping and manipulation technologies.

\begin{figure}[htbp]
\centering
\includegraphics[width=0.45\textwidth]{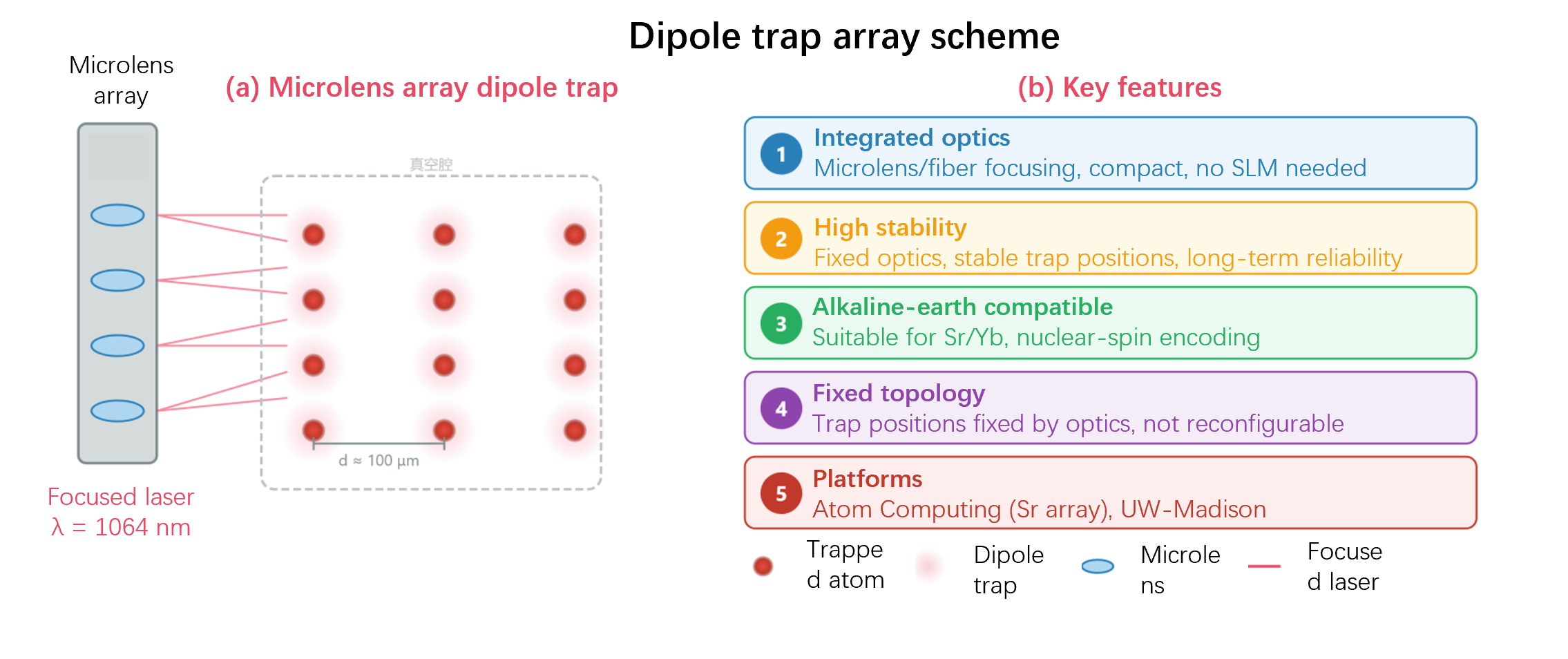}
\caption{Schematic of dipole trap array (microlens array and integrated photonics). (a) Microlens array scheme. (b) Integrated photonic chip scheme (adapted from Refs. [12,51].)}
\label{fig:6}
\end{figure}

\subsection{Key Enabling Technologies}

The development of neutral-atom quantum computing depends on advances in several key enabling technologies:

\textbf{Rydberg-state preparation:} High-efficiency, low-decoherence Rydberg excitation is a prerequisite for achieving high-fidelity quantum gates. Two-photon excitation schemes can increase Rydberg excitation efficiency to over 95\% by selecting appropriate intermediate levels and laser parameters. Optimal control techniques (such as the GRAPE algorithm) further reduce leakage errors during excitation by optimizing pulse waveforms.

\textbf{Non-destructive readout:} Quantum computing requires reading out qubit states without destroying quantum information. Fluorescence imaging is the most commonly used readout method: atoms are illuminated with resonant light to emit fluorescence, which is detected by high-numerical-aperture lenses and EMCCD cameras. For mid-circuit measurement, it is necessary to use the light-shift differences of different atomic states to separate the measurement qubit from data qubits in frequency, enabling selective measurement of the target qubit without disturbing other qubits.

\textbf{Quantum error correction codes:} The reconfigurable connectivity of the neutral-atom platform makes it naturally suitable for implementing various quantum error correction codes. The surface code uses nearest-neighbor connections to achieve fault tolerance and is the most widely studied scheme. In recent years, leveraging the all-to-all connectivity of neutral atoms, important progress has also been made in low-density parity-check (LDPC) codes, Bacon--Shor codes, and transversal gate schemes. In 2024, Bluvstein et al. implemented a logical quantum processor based on transversal gates; in 2025, Pecorari et al. proposed high-rate LDPC codes for neutral-atom registers, and Bravyi et al. proposed quantum codes with high thresholds and low overhead.

\textbf{Atom replenishment:} Because of the finite trap lifetime of atoms (typically several seconds to tens of seconds), atom loss inevitably occurs during long-duration quantum computation. Atom replenishment technology prepares spare atoms in a storage array and replenishes them in real time when atom loss is detected, enabling continuous-operation quantum computing. This technology is a necessary condition for achieving large-scale fault-tolerant quantum computing.

\begin{table}[htbp]
\centering
\small
\caption{Quantitative comparison of mainstream technical routes}
\label{tab:1}
\begin{tabular}{p{1.8cm} p{2.0cm} p{1.6cm} p{1.6cm}}
\toprule
\textbf{Metric} & \textbf{Tweezer + Rydberg} & \textbf{Lattice} & \textbf{Dipole} \\
\midrule
Max scale & $\sim$11,000 [63]; 6,100 [40] & $\sim$1,000+ [6] & $\sim$hundreds \\
Reconfig. time & $\sim$60 ms [2] & Non-reconfig. & Limited \\
Max 2Q fidelity & 99.84\% [2] & $\sim$97--98\% [6] & $\sim$95--97\% \\
$T_2$ coherence & 300 ms (Rb); 12.6 s (Cs [40]) & Several s & Several s \\
Trap lifetime & 6,000 s [24] & $\sim$60 s & $\sim$10--60 s \\
Addressing & Strong & Weak & Medium \\
Connectivity & All-to-all & Nearest & Medium \\
Platforms & QuEra, Atom, China & Pasqal & Research \\
Advantages & Flexible, high-fidelity & Parallel sim. & Compact \\
Bottlenecks & Laser, loss & Addressing, fidelity & Scalability \\
\bottomrule
\end{tabular}
\end{table}

\section{Domestic and International Research Progress}

The development of neutral-atom quantum computing can be divided into three stages: the theoretical foundation period (2000--2010), the technology development period (2010--2020), and the explosive growth period (2021--2026). Figure 7 summarizes the key milestones of each stage.

\begin{figure}[htbp]
\centering
\includegraphics[width=0.45\textwidth]{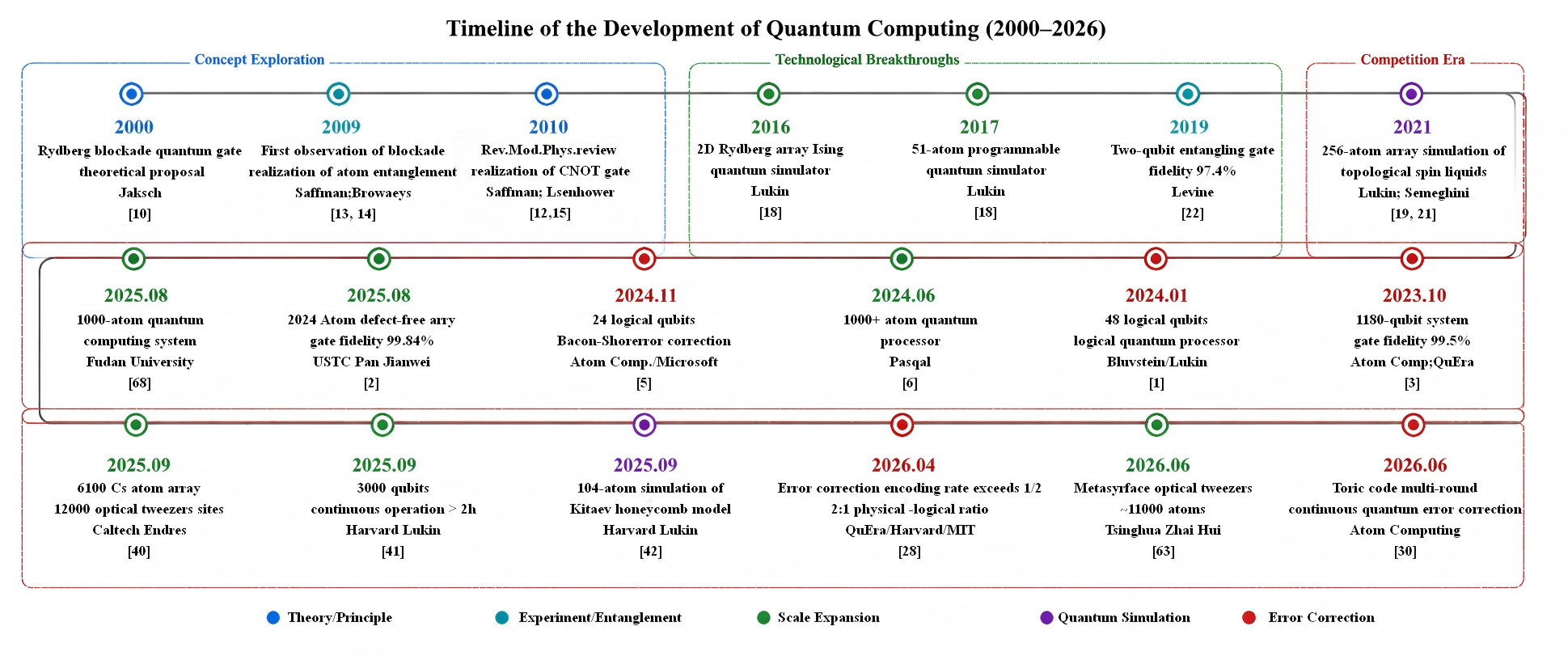}
\caption{Key milestones in neutral atom quantum computing (2000--2026). A serpentine timeline showing 19 landmark events, divided into three stages: theoretical foundation (blue), technology development (green), and explosive growth (red). Node colors indicate event categories: theory/principle (blue), experiment/entanglement (cyan), scale expansion (green), quantum simulation (purple), and error correction (red).}
\label{fig:7}
\end{figure}

\subsection{Early Foundations (2000--2010)}

The theoretical foundation of neutral-atom quantum computing dates back to 2000. Jaksch et al. first proposed using the strong dipole interactions between Rydberg atoms to implement fast quantum gates, demonstrating that the Rydberg blockade mechanism could be used to deterministically entangle neutral atoms. This theoretical scheme opened the research direction of neutral-atom quantum computing. In 2005, Saffman and Walker performed a systematic feasibility analysis and device design of the scheme, quantifying the main decoherence mechanisms limiting gate fidelity---spontaneous emission from Rydberg states, Doppler decoherence caused by atomic thermal motion, laser phase noise, coupling of motional states in the trap, and leakage errors due to incomplete blockade---and demonstrating the feasibility of implementing single-qubit and two-qubit gates at MHz rates with per-operation fidelity errors controlled at the $10^{-3}$ level, establishing quantitative design benchmarks for subsequent experimental schemes.

In 2010, Saffman, Walker, and Mølmer published a review article entitled ``Quantum information with Rydberg atoms'' in \textit{Reviews of Modern Physics}, systematically expounding the physical principles, experimental techniques, and application prospects of quantum information processing with Rydberg atoms, and becoming a foundational reference for the field. This review discussed in detail the physical mechanism of Rydberg blockade, quantum gate implementation schemes, sources of decoherence, and scaling paths, laying a theoretical framework for research in the following decade.

On the experimental side, in 2009 two research groups almost simultaneously achieved the first experimental verification of Rydberg blockade. The Saffman and Walker group at the University of Wisconsin (Urban et al.) and the Browaeys group at the Institut d'Optique in France (Gaëtan et al.) independently observed the Rydberg blockade effect between two trapped atoms and realized atomic entanglement. Although the entanglement fidelities of these experiments were limited (Bell-state fidelities of about 0.58--0.71), they demonstrated for the first time in principle the feasibility of using Rydberg blockade to entangle neutral atoms. In 2010, Isenhower et al. used Rydberg blockade to implement a CNOT gate operation, marking the transition of neutral-atom quantum gates from conceptual verification to actual operation.

\subsection{Technology Development Period (2010--2020)}

The period from 2010 to 2020 was a key development stage for neutral-atom quantum computing, from proof of principle to hundred-atom-scale systems. Progress in this stage was mainly reflected in three aspects: the maturation of optical tweezer array technology, the continuous improvement of quantum gate fidelity, and the rapid expansion of quantum simulation scale.

\textbf{Maturation of optical tweezer array technology:} Around 2016, the Endres team and the Browaeys team respectively achieved the technology of trapping and rearranging individual atoms using optical tweezer arrays, making the preparation of defect-free atomic arrays possible. In 2016, Labuhn et al. reported in \textit{Nature} tunable two-dimensional arrays of Rydberg atoms and realized Ising-type quantum simulation. In 2018, Barredo et al. reported in \textit{Nature} three-dimensional atomic structures assembled atom by atom, expanding neutral-atom arrays from two dimensions to three dimensions and opening new spatial dimensions for scaling.

\textbf{Expansion of quantum simulation scale:} In 2017, Bernien et al. (Lukin group) reported in \textit{Nature} a 51-atom quantum simulator that used Rydberg interactions to simulate quantum many-body dynamics. This was the first programmable quantum simulator with more than 50 atoms, verifying the capability of the neutral-atom platform in quantum simulation.

\textbf{Improvement of quantum gate fidelity:} During this period, the fidelity of Rydberg blockade quantum gates continued to increase. In 2019, Levine et al. reported two-qubit entangling gates with a fidelity of 97.4\%.

\subsection{Explosive Growth Period (2021--2026)}

Since 2021, neutral-atom quantum computing has entered a period of explosive development, with many milestone achievements emerging one after another. This section reviews progress in four aspects: ``physical basis---hardware scale---error correction and fault tolerance---application exploration.''

\paragraph{(1) Physical Basis}

\textbf{Quantum gate fidelity and mid-circuit measurement technology.} In 2022, Graham et al. reported in \textit{Nature} multi-qubit entanglement operations. At the same time, mid-circuit measurement technology also made breakthroughs: Deist et al. (2022) and Graham et al. (2023) respectively demonstrated measurement of auxiliary qubits without destroying data qubits, which is a technical prerequisite for quantum error correction. In 2026, improvements in gate fidelity and readout technology continued to advance: Stein et al. proposed a multi-target Rydberg gate scheme based on spatial blockade engineering, and Li et al. proposed a method using auxiliary drives to accelerate Rydberg entangling gates; these works improve parallelism and fidelity from the perspective of gate operation design. Tsai et al. proposed a gate-based readout and cooling scheme using auxiliary qubits, demonstrating the use of Rydberg entangling gates and auxiliary atoms to achieve multiple rounds of non-destructive readout and conversion of quantum information from electronic states to motional states, applicable to mid-circuit measurement and atomic cooling. In May 2026, Infleqtion reported a record rubidium--cesium two-species entangling gate, while Norrell and Saffman et al. proposed a theoretical path toward neutral-atom entangling gate fidelities above 99.9\%, analyzing the main factors limiting fidelity improvement (spontaneous emission from Rydberg states, laser phase noise, atomic thermal motion, etc.) and corresponding improvement strategies, pointing the way to breaking through the current 99.5\% gate fidelity bottleneck.

In terms of innovations in gate operation schemes, domestic research teams have proposed a Rydberg blockade CZ gate scheme based on off-resonant modulated driving (ORMD). In 2020, Sun et al. proposed a protocol for implementing a CZ gate using specially designed smooth-waveform pulses; this scheme requires only a single modulated pulse to complete the gate operation, needs no independent addressing, and is insensitive to the precise value of the blockade shift, effectively suppressing population leakage errors and rotation errors. In 2022, Fu et al. implemented this scheme in a rubidium atom experiment, achieving a CZ gate through a single-modulated-pulse-driven two-photon transition and generating a two-atom entangled state combined with global microwave pulses, with a raw fidelity of 94.5\%. In 2023, Sun further proposed a systematic method to suppress high-frequency components in the modulation waveform, filtering out high-frequency components while maintaining two-qubit and multi-qubit gate conditions, improving gate fidelity, operation speed, and robustness, and could be combined with double-pulse techniques to overcome the effects of atomic thermal motion on gate fidelity. This technical route provides a new scheme from the perspective of waveform engineering for improving two-qubit gate fidelity in large-scale arrays.

\textbf{Improvement of trap lifetime.} In 2021, Schymik et al. reported in \textit{Physical Review Applied} an experiment achieving a 6000-second trap lifetime in a cryogenic environment, greatly extending the usable time of atoms and making long-duration quantum computation possible.

\paragraph{(2) Hardware Scale}

\textbf{Continuous expansion of quantum simulation scale.} In 2021, Ebadi et al. expanded the scale to 256 atoms, Scholl et al. (Browaeys group) used 196 Rydberg atoms to simulate two-dimensional antiferromagnets, and Semeghini et al. used 219 atoms to realize simulation of topological quantum spin liquids. Since then, the scale and depth of quantum simulation have continued to advance: in 2025, Evered et al. used 104 atoms to simulate the Kitaev honeycomb model, and in 2026, White et al. implemented quantum cellular automata on a dual-species Rydberg processor (see subsection (4)). These works show that the neutral-atom platform has already possessed large-scale quantum simulation capabilities and continues to lead in simulating complex quantum many-body physics.

\textbf{Thousand-qubit scale and ten-thousand-atom array breakthroughs.} In October 2023, Atom Computing announced that its second-generation system achieved more than 1180 neutral-atom qubits, surpassing the superconducting quantum computing platform in physical qubit count for the first time. In June 2024, Pasqal announced a quantum processor with more than 1000 atoms in a single loading. In August 2025, Pan Jianwei's team at the University of Science and Technology of China used artificial-intelligence-driven parallel rearrangement technology to construct defect-free two-dimensional and three-dimensional arrays of 2024 atoms within 60 milliseconds, setting a new world record. The single-qubit gate fidelity of this system reached 99.97\%, the two-qubit gate fidelity reached 99.84\%, and the detection fidelity reached 99.92\%, reaching an internationally leading level in key indicators. This achievement was selected by the American Physical Society's \textit{Physics} magazine as one of the nine major international physics advances of 2025.

In terms of array scale, in September 2025 the Endres group at the California Institute of Technology reported a qubit array with 6100 cesium atoms trapped in 12,000 optical tweezer sites, with single-qubit manipulation accuracy reaching 99.98\%, and demonstrated coherent atom transport. In the same period, the Lukin group at Harvard University and collaborators including Hollerith at ETH Zurich reported in \textit{Nature} the continuous operation of a 3000-qubit coherent system. This work used a dual optical-lattice conveyor belt to continuously transport atoms from a reservoir to the computing zone, repeatedly loading tweezers at a rate of 300,000 atoms per second, achieving continuous operation of an array with more than 3000 atoms for more than two hours, far exceeding the 60-second trap lifetime of optical tweezers. The experiment further demonstrated that fresh atoms could be continuously replenished while maintaining the coherence of already stored qubits, opening a feasible engineering path to solving the core bottleneck of atom loss. In June 2026, Zhai Hui's team at Tsinghua University reported on arXiv an experiment using a single-piece metasurface optical tweezer array to simultaneously trap about 11,000 neutral atoms, breaking the ten-thousand-atom threshold for neutral-atom arrays for the first time and also marking the first time that usable qubit resources in any quantum computing platform reached the ten-thousand level. The scheme uses a metasurface about 2 cm in diameter placed outside the vacuum chamber; without needing a microscope objective, it can generate the entire optical tweezer array, maximizing laser power efficiency. In a $135 \times 135$ array (18,225 sites in total), it achieved stable trapping of an average of 11,022 atoms (filling rate 60.5\%, standard deviation of atom number between experiments only 167). The team developed the ``Zhui Feng'' rearrangement algorithm combining graph neural networks with an efficient parallel decoder, compressing the fast rearrangement time of ten-thousand-atom-scale arrays to within 20 milliseconds.

\textbf{Array control and hardware technology.} High-speed dynamic control of large-scale arrays is a key enabling technology for scaling. In 2026, Wei et al. reported a 10 MHz refresh-rate spatial light modulator, increasing the control bandwidth of optical tweezers by two orders of magnitude and serving as a core device for high-speed dynamic rearrangement of large-scale arrays. Bytyqi et al. (Vuletić group) developed a MHz-rate two-dimensional optical potential scanning device, providing a new tool for high-speed atom manipulation. These hardware technology breakthroughs help alleviate the bandwidth bottleneck of large-scale array control, but developing application-specific integrated circuits (ASICs) or using time-multiplexing techniques to achieve fully parallel control above ten thousand qubits still requires substantial engineering investment.

\paragraph{(3) Systematic Advancement of Quantum Error Correction and Fault-Tolerant Computing}

\textbf{Logical qubits and quantum error correction.} In December 2023, the Lukin group at Harvard University, in collaboration with QuEra, reported the first programmable logical quantum processor. The system used 280 rubidium atoms to realize encoding of 48 logical qubits and executed fault-tolerant quantum algorithms in a three-zone architecture (storage, entanglement, and readout zones) through transversal gate operations. Experiments showed that when running quantum algorithms with hundreds of logical gates using logical qubits, performance was up to ten times higher than when using physical qubits, marking the transition of fault-tolerant quantum computing from theory to practice.

In November 2024, Atom Computing and Microsoft used the Bacon--Shor code to create and entangle 24 logical qubits on more than 1200 physical qubits; 28 logical qubits ran the Bernstein--Vazirani algorithm with higher accuracy than unencoded physical qubits, achieving active quantum error correction on a commercial platform for the first time.

In 2026, quantum error correction achieved multiple landmark advances. In April, QuEra, Harvard University, and MIT reported a quantum error correction encoding rate breakthrough exceeding 1/2---based on Kasai's theoretical breakthrough, it was proved that only no more than two physical qubits are needed to construct a reliable logical qubit. It should be specifically noted that this 2:1 encoding rate was verified only for memory (storage), i.e., maintaining quantum states without executing logical gate operations; the encoding rate in error-correction computation scenarios still requires further research. Simulated verification showed error rates as low as once per trillion steps, far better than the hundreds to thousands of physical qubits overhead required by traditional surface codes. In June, Atom Computing announced the industry's first complete quantum error correction demonstration based on the toric code, proving that the logical error rate of its neutral-atom system decreases as the number of physical qubits increases, becoming the second company after Google to achieve multi-round continuous quantum error correction and the first to reach this milestone on a neutral-atom platform. The toric code and surface code belong to the same family of topological codes, both using redundant encoding on two-dimensional lattices to achieve fault tolerance; the main difference is that the toric code is defined on a torus lattice, and its logical operators correspond to non-local winding operators, while the surface code can be viewed as a variant of the toric code on a planar sheet with boundaries, introducing additional logical qubits through boundary conditions. In terms of decoding algorithms, both use matching decoders (such as minimum-weight perfect matching), but the symmetric structure of the toric code can provide better threshold performance under some error models.

\textbf{Fault-tolerant architecture innovations.} In 2024, Xu et al. reported in \textit{Nature Physics} a fault-tolerant quantum computation scheme with constant overhead, which is expected to reduce the space overhead of quantum error correction. In September 2025, QuEra, Harvard University, and Yale University published the ``Algorithmic Fault Tolerance'' (AFT) framework in \textit{Nature}, reducing runtime overhead by more than 30 times through transversal operations and correlated decoding. In the same period, the proposal of high-rate LDPC codes for neutral-atom registers, and high-threshold low-overhead quantum codes enriched the fault-tolerant scheme options for the neutral-atom platform. Ismail et al. further proposed the transversal STAR architecture, combining transversal Clifford operations with high-rate qLDPC codes and using small-angle magic-state injection to achieve effective quantum simulation, bridging the gap between near-term analog applications and fully fault-tolerant computation. In 2026, Bluvstein et al. published in \textit{Nature} a fault-tolerant neutral-atom architecture for universal quantum computing, systematically demonstrating the path to fault-tolerant quantum computing using reconfigurable atom arrays, including new technologies such as loss-aware decoding, providing a systematic solution for achieving fault-tolerant computation in the presence of atom loss.

\textbf{Logical-level quantum algorithm demonstrations.} In July 2025, QuEra, Harvard University, and MIT demonstrated logical-level magic-state distillation on a neutral-atom quantum computer, a key step toward universal fault-tolerant quantum computing. This result shows that the neutral-atom platform already has the ability to execute advanced fault-tolerant quantum algorithm primitives.

In September 2025, Rines et al. (Infleqtion team, with Saffman as a collaborator) demonstrated a logical qubit architecture on Infleqtion's Sqale quantum processor, integrating atomic motion and in-place entanglement operations. This work achieved three key results: for the first time, a precompiled version of Shor's algorithm was executed on logical qubits, and logical performance was better than physical qubits after combining loss correction and leakage detection; a constant-depth logical CNOT ladder circuit was constructed, achieving 2--4 times error reduction on 8 and 12 logical qubits; and a many-hypercube code was prepared and single-round decoding post-processing error correction was executed, with logical-state performance eight times better than the physical state. This is the first demonstration of Shor's algorithm on a neutral-atom platform using logical qubits, marking the transition of the neutral-atom platform from logical qubit verification to logical-level quantum algorithm execution.

\paragraph{(4) Application Exploration}

\textbf{Frontier exploration of quantum simulation.} In September 2025, Evered et al. (Lukin group) reported in \textit{Nature} a quantum simulation of the Kitaev honeycomb model using 104 neutral-atom qubits. Through Floquet engineering, three-body interactions between atoms were realized in an optical tweezer array, and a topological spin liquid state was successfully prepared and verified, marking the first simulation of this classic problem in condensed matter physics on a quantum computer. This result shows that the neutral-atom platform is also a powerful tool for exploring complex quantum many-body physics beyond quantum computing. In 2026, White et al. reported the implementation of quantum cellular automata on a dual-species Rydberg processor, revealing a new paradigm for dual-atom-species platforms in quantum information processing.

\textbf{Near-term quantum application exploration.} In addition to quantum simulation and fault-tolerant computing, the neutral-atom platform also has application potential in quantum machine learning and variational quantum algorithms. In December 2024, Beaulieu et al. (QuEra in collaboration with Merck, Amgen, and Deloitte) reported a molecular property prediction method based on neutral-atom quantum reservoir computing, demonstrating the generalization advantage of quantum methods over traditional machine learning models in small-data-set scenarios (100--300 samples), providing empirical evidence for near-term applications of quantum computing in drug discovery and other fields. In September 2025, Matwiejew et al. experimentally demonstrated for the first time on neutral-atom hardware a variational ansatz based on continuous-time quantum walk, deriving analytical expressions for near-optimal control parameters for non-entangled target states that can be directly ported to hardware with minimal calibration, providing a feasible path for deploying variational quantum algorithms on analog-mode neutral-atom platforms.

\section{Domestic and International Industrialization Progress}

In recent years, neutral-atom quantum computing has rapidly moved from laboratory research to industrialization. Many start-ups and research institutions have launched commercial quantum computing platforms or prototypes. This section reviews the international industrial landscape and domestic progress separately.

\subsection{International Industrial Landscape}

The industrialization of neutral-atom quantum computing abroad is mainly driven by the following companies:

\textbf{QuEra Computing (USA):} Founded in 2018 in Boston and incubated by the Lukin group at Harvard University, QuEra is a pioneer in neutral-atom quantum computing. Its ``Aquila'' quantum processor was launched on the Amazon Braket cloud platform in 2022, becoming the first publicly available neutral-atom quantum computer. In February 2025, QuEra completed a US\$230 million financing round to accelerate the development of large-scale fault-tolerant quantum computers. In September 2025, QuEra, together with Harvard University and Yale University, published the Algorithmic Fault Tolerance (AFT) framework in \textit{Nature}. In April 2026, QuEra, together with Harvard University and MIT, reported a quantum error correction scheme with an encoding rate breakthrough exceeding 1/2, reducing the ratio of physical qubits to logical qubits from hundreds-to-one to about 2:1 (for memory functions), greatly reducing the physical resource requirements for fault-tolerant quantum computing. QuEra's roadmap plans to use 10,000 atoms to generate 100 logical qubits by 2026. In March 2025, QuEra joined the NVIDIA Accelerated Quantum Research Center (NVAQC) as a founding member; in April it was selected for the first phase of the US DARPA Quantum Benchmarking program.

\textbf{Atom Computing (USA):} Founded in 2018 in Berkeley, California, by Ben Bloom and Jonathan King. Atom Computing adopts the alkaline-earth-metal strontium atom scheme, using nuclear-spin-state encoding to achieve extremely long coherence times. Its second-generation system achieved 1180+ physical qubits in 2023, and the third-generation Magne system is planned for delivery in 2027 with 50 logical qubits. Atom Computing works deeply with Microsoft Azure Quantum, and its system has been integrated into the Azure Quantum Elements platform. In 2025, it was named by \textit{Fast Company} as the 10th most innovative company in the world. In June 2026, Atom Computing announced the industry's first complete quantum error correction demonstration based on the toric code, proving that the logical error rate of its neutral-atom system decreases as system scale increases, becoming the first company to achieve multi-round continuous quantum error correction on a neutral-atom platform.

\textbf{Pasqal (France):} Founded in 2019 in Paris and incubated by the Browaeys and Lahaye teams at the Institut d'Optique. Pasqal adopts the rubidium atom and optical tweezer/optical lattice hybrid scheme, emphasizing analog--digital hybrid quantum computing. Between 2024 and 2025, Pasqal deployed its 100+ qubit Orion Beta system at the French GENCI and German Jülich supercomputing centers, becoming one of the first quantum computers deployed on-site in supercomputing facilities. In February 2026, Pasqal delivered a 140+ qubit system to the Italian CINECA supercomputing center and launched the Vela QPU (256+ high-quality qubits), further expanding its on-site deployment scale. Its roadmap plans to achieve 10,000 qubits by 2028 and 100 logical qubits by 2029.

\textbf{Infleqtion (USA):} Formerly known as ColdQuanta, Infleqtion is an important participant in the neutral-atom quantum computing field. In May 2026, Infleqtion reported multiple technical breakthroughs, including a record rubidium--cesium two-species entangling gate, 17 $\mu$K sub-Doppler cooling temperature, 17 cm light transmission, and a theoretical path toward entangling gate fidelities above 99.9\%. These technologies support a continuous-operation architecture in which the atom preparation region is spatially separated from the computing region. In September 2025, the Infleqtion team demonstrated a logical qubit architecture on its Sqale quantum processor and executed Shor's algorithm on logical qubits for the first time.

Oratomic (France/USA): Founded in 2022 and spun out from the groups of Antoine Browaeys and Thierry Lahaye (the same origin as Pasqal), Oratomic focuses on high-density 3D optical tweezer arrays and has recently demonstrated record-breaking atomic array densities and fast parallel manipulation. In 2025, Oratomic announced a 1000+ atom array with sub-$\mu$m spacing, offering a unique path for compact scalable quantum processors. The company is actively developing mid-scale quantum simulators and collaborating with European high-performance computing centers.

In addition, teams related to the University of Toronto in Canada, multiple European research institutions, and several Japanese research groups are actively advancing the research and development of neutral-atom quantum computing.

\textbf{Roadmap commentary:} The roadmap goals of the above companies are generally quite aggressive---for example, QuEra plans to achieve 100 logical qubits by 2026, and Pasqal plans to achieve the same goal by 2029. However, these timelines face multiple risks: (1) The leap from the current 2:1 memory encoding rate to error-correction computation has not yet been experimentally verified, and the error-correction overhead in logical gate operations may be significantly higher than in memory scenarios---taking the surface code as an example, error-correction computation requires additional transversal gates and magic-state distillation, and the space--time overhead may increase to 5--10 times that of memory scenarios; (2) roadmaps generally assume that gate fidelity can be improved from the current 99.5\% level to above 99.9\%, but theoretical analysis by Saffman et al. shows that the main factors currently limiting fidelity include spontaneous emission from Rydberg states (contributing about 0.05\%--0.1\% error rate), laser phase noise (contributing about 0.1\%--0.2\%), and interaction fluctuations caused by atomic thermal motion (contributing about 0.1\%), with the sum of the three already approaching the physical limit of 99.9\% fidelity, and breaking through this barrier may require new physical mechanisms such as F\"{o}rster resonance gates; (3) the engineering complexity of atom replenishment and real-time error-correction decoding required by continuous-operation architectures is often underestimated in roadmaps---real-time decoders need to complete syndrome data processing on the microsecond timescale, placing extremely high demands on the decoding throughput of FPGAs/ASICs. Therefore, these timelines are better understood as optimistic estimates of technical goals rather than engineering commitments.

\subsection{Domestic Progress}

China started relatively late in neutral-atom quantum computing but has developed rapidly in recent years, forming a pattern of coordinated advancement by academia and industry.

\textbf{Academia:} Pan Jianwei's team at the University of Science and Technology of China has achieved multiple representative results in neutral-atom quantum computing. In August 2025, the team used artificial intelligence to drive a high-speed spatial light modulator to realize an array-scale-independent constant-time atom rearrangement scheme, constructing defect-free two-dimensional and three-dimensional arrays of 2024 atoms within 60 milliseconds; the related results were published in \textit{Physical Review Letters} as an ``Editor's Suggestion.'' The single-qubit gate fidelity of this system reached 99.97\%, the two-qubit gate fidelity reached 99.84\%, and the detection fidelity reached 99.92\%. For comparison, in 2023 the Harvard University/QuEra team achieved a two-qubit gate fidelity of 99.5\% at a scale of 60 atoms, and in 2025 the California Institute of Technology reported a single-qubit manipulation accuracy of 99.98\%. It can be seen that in key indicators such as gate fidelity and detection fidelity, China has reached an internationally leading level. This achievement was selected by the American Physical Society's \textit{Physics} magazine as one of the nine major international physics advances of 2025. In 2026, the team further reported an experimental scheme for sustaining high-fidelity quantum logic in neutral-atom circuits via mid-circuit operations (arXiv:2603.01612), demonstrating the ability to perform intermediate measurements and feedback control without destroying data qubits, which is a key technical prerequisite for realizing quantum error correction cycles.

Multiple domestic research teams have made important progress in rearrangement algorithms for neutral-atom arrays, optical metasurface tweezer technology, and gauge-theory quantum simulation. Wang et al. proposed a maximum-parallelism algorithm to accelerate the assembly of defect-free atomic arrays, improving the construction efficiency of large-scale arrays. Zhang et al. used a single-piece metasurface to directly generate a large-scale array containing 78,400 tweezers, providing a new scheme for the compactness and scalability of tweezer systems. On this basis, Wang et al. (Zhai Hui's team at Tsinghua University) used a single-piece metasurface optical tweezer array to achieve stable trapping of about 11,000 atoms, breaking the ten-thousand-atom threshold for neutral-atom arrays for the first time. Cheng and Zhai systematically elaborated from a theoretical perspective the methods and prospects of using Rydberg atom arrays to simulate lattice gauge theories.

The Innovation Academy for Precision Measurement Science and Technology, Chinese Academy of Sciences, has deep accumulation in neutral-atom qubit manipulation. Yang et al. proposed and experimentally verified the ``magic-intensity optical trap'' scheme, effectively maintaining the coherence of single-atom qubits during transfer between optical traps. Zeng et al. used Rydberg blockade to realize deterministic entanglement between two different isotopes of neutral atoms ($^{87}$Rb and $^{85}$Rb), demonstrating heteronuclear CNOT gate operations. In November 2025, Li et al. from the team reported in \textit{Nature Communications} a neutral-atom quantum computing architecture based on a fiber array, using microlenses on fiber end faces to achieve independent addressing and manipulation of atoms, providing a compact and scalable new scheme for parallel control of large-scale atom arrays.

The neutral-atom quantum computing team of the Department of Physics at Fudan University has made systematic progress in the development of neutral-atom quantum computing systems and quantum solving of NP problems. Since 2020, the team has proposed a quantum annealing architecture based on quantum wires, providing a scheme for solving general NP computational problems with neutral atoms. In 2023, through atom--cavity coupling technology, the encoding complexity was further reduced to the theoretical limit, making the experimental solution of large-scale NP problems feasible. Recently, based on parallel moving technology with ultra-low atom loss rate (below 1\%), a quantum computing parallel instruction set was formulated, a highly streamlined quantum black box was constructed that reduces circuit depth from $O(\sqrt{n})$ to $O(\log^2 n)$, achieving exponential efficiency improvement. In August 2025, the team developed a neutral-atom quantum computing system containing 1000 atoms and developed an AI-assisted quantum control scheme for automatic calibration and optimization of large-scale atomic systems; the self-developed objective lens performance can already support a ten-thousand-qubit-scale neutral-atom quantum computing platform.

\textbf{Industry:} Zhongke Kuyuan Technology (Wuhan) Co., Ltd. developed China's first engineering neutral-atom quantum computer, ``Hanyuan 1,'' in 2024, filling the gap in China's neutral-atom quantum computing technology route, and successfully delivered two neutral-atom quantum computing experimental platforms based on the Hanyuan 1 architecture in early 2025, realizing China's first commercial delivery of neutral-atom quantum computers. In 2026, the company launched ``Hanyuan 2,'' the world's first dual-core neutral-atom quantum computer. The system adopts a dual-cavity structure design, with more than 500 optical tweezers, a coherence time ($T_2$) exceeding 300 ms, single-qubit gate fidelity better than 99.9\%, and two-qubit gate fidelity better than 99\% (measured by randomized benchmarking methods); the localization rate of equipment exceeds 80\%. The standard cabinet design keeps the overall power consumption below 7 kW, and because no extremely low-temperature environment is required, it can be deployed in ordinary machine rooms. The company plans to increase the number of qubits to 1000 in the next-generation ``Hanyuan 3.''

Taiyi Liangsheng (Shanghai) Quantum Technology Co., Ltd. was established in January 2026 and is China's first full-system company focusing on ytterbium (Yb) neutral-atom quantum computing. Different from the mainstream rubidium/cesium schemes, ytterbium atoms have richer energy-level structures, enabling faster qubit gate operations, native high-precision multi-qubit gates, and multi-photon interconnection in the communication band. The company has completed angel-round financing of more than 100 million yuan and Pre-A round financing of 300 million yuan, and plans to demonstrate logical qubits through error correction by the end of 2026.

Liangyi Wanxiang ( China): Founded in 2024 and headquartered in Beijing, Liangyi Wanxiang is a dedicated neutral-atom quantum computing full-system company. Its core team has deep expertise in Rydberg quantum gates, high-speed feedback control, and integrated photonics. In 2025, the company demonstrated a 500+ atom tweezer array with a two-qubit gate fidelity above 99.5\% and announced plans to build a 1000+ qubit error-corrected platform by 2027. Liangyi Wanxiang has also established collaborations with domestic foundries for custom optical chips, aiming at compact and scalable quantum processors.

Zhongqi Wuliang is a domestic neutral-atom quantum computing full-system company; its founding team members participated in the development of what was then the world's largest neutral-atom quantum computer, with related results published in \textit{Nature}. In 2026, Zhongqi Wuliang and Xuanxiang Technology, a metasurface optical chip company, jointly completed the world's first neutral-atom chip-level million-tweezer verification. Xuanxiang Technology released the world's first metasurface chip capable of generating million-level atom tweezer arrays, breaking through the scaling bottlenecks of traditional spatial light modulators and acousto-optic deflectors in pixel size and optical aperture; Zhongqi Wuliang provided the neutral-atom experimental platform and completed system adaptation and optical-field verification. This scheme has the industrialization advantages of standardized integration and batch iteration. The two companies plan to achieve 100,000-level atom loading and stable capture in the medium term and move toward the million-atom scale in the long term.

In addition, multiple domestic start-ups (such as Buchou Quantum and Yuanzi Xinguang) are actively laying out the neutral-atom quantum computing industry chain.

According to the Photon Box Research Institute report ``2026 Global Quantum Computing Industry Development Outlook,'' China's core technologies and industrial ecosystem in quantum computing are accelerating toward independent control. Multiple domestic enterprises and research institutions are actively laying out the neutral-atom quantum computing industry chain, from lasers, optical components, and vacuum systems to measurement-and-control electronic equipment and software stacks, gradually building a complete technical system.

Reviewing the explosive progress of the 2021--2026 period, neutral-atom quantum computing shows the following trends: (1) the speed of scaling far exceeds previous expectations---from arrays of tens of atoms in 2021 to defect-free arrays of 2024 atoms in 2025, with an average annual growth rate of more than 100\%; (2) gate fidelity continues to improve but is approaching the current technical limit---two-qubit gate fidelity has increased from the 97\% level to 99.84\%, and further breakthrough above 99.9\% will require new physical mechanisms (such as F\"{o}rster resonance gates); (3) quantum error correction is moving from proof of principle to engineering implementation---from logical qubit demonstrations to toric code multi-round error correction and 2:1 memory encoding rates, but error-correction computation (rather than only error-correction memory) remains an unresolved core challenge; (4) the industrialization process is accelerating but the business model has not yet been verified---multiple companies have released roadmaps, but the transition from laboratory prototypes to commercially available fault-tolerant quantum computers still requires solving systemic problems in engineering, reliability, and cost.

\section{Development Bottlenecks and Challenges}

Although neutral-atom quantum computing has made significant progress, moving from the current stage of hundreds to thousands of physical qubits to million-qubit-scale practical fault-tolerant quantum computing still faces multiple core bottlenecks.

\subsection{Scalability--Fidelity Trade-off}

The scalability advantage of the neutral-atom platform in terms of qubit number has been fully verified, but there is an inherent contradiction between increasing qubit number and maintaining gate operation fidelity. As array scale expands, problems such as crosstalk between atoms, light-field inhomogeneity, and trap frequency differences gradually become prominent. The current two-qubit gate fidelity of 99.5\% has exceeded the theoretical threshold of the surface code (about 99\%), but the threshold theorem requires all operations in the error-correction circuit to satisfy the threshold conditions; actual error-correction circuits must also consider measurement errors and atom loss, and there is still a gap from the above 99.9\% fidelity required for low-overhead fault tolerance. In 2026, Saffman et al. proposed a theoretical path toward neutral-atom entangling gate fidelities above 99.9\%, analyzing the main factors limiting fidelity improvement (spontaneous emission from Rydberg states, laser phase noise, atomic thermal motion, etc.) and corresponding improvement strategies. Li et al.'s auxiliary-drive acceleration scheme and Stein et al.'s multi-target Rydberg gate scheme provide new paths for improving parallelism and fidelity from the perspective of gate operation design. Higher gate fidelity means fewer physical qubit overheads---taking the surface code as an example, improving gate fidelity from 99\% to 99.9\% can reduce the number of physical qubits required per logical qubit by an order of magnitude. Therefore, improving gate fidelity in large-scale arrays remains one of the most urgent technical challenges.

\subsection{Engineering Implementation of Quantum Error Correction}

Although logical qubits and quantum error correction have been verified in principle experimentally, expanding from tens of logical qubits to thousands or even tens of thousands of logical qubits still faces enormous challenges. Quantum error correction requires syndrome measurement and decoding after each logical gate operation, which demands extremely high measurement efficiency and real-time feedback control. Current mid-circuit measurement technology has been demonstrated, but its fidelity and speed are not yet sufficient to support large-scale error-correction cycles. In 2026, multiple important advances were achieved: Atom Computing realized multi-round continuous quantum error correction based on the toric code, proving the key criterion that logical error rate decreases as system scale increases; the QuEra team reduced the ratio of physical qubits to logical qubits to about 2:1 (for memory functions); Bluvstein et al. proposed a fault-tolerant architecture including loss-aware decoding, providing a systematic solution for achieving fault-tolerant computation in the presence of atom loss. However, the leap from error-correction memory (memory) to error-correction computation (computation)---that is, executing logical gate operations under error-correction protection rather than only maintaining quantum states---still requires substantial work. Taking the surface code as an example, when the physical gate fidelity is 99\%, about 100--200 physical qubits are needed to encode one logical qubit (code distance 7--9), and when fidelity is improved to 99.9\%, this can be reduced to 20--50 (code distance 3--5). The 2:1 encoding rate reported by QuEra used LDPC codes rather than surface codes and was verified only for memory functions; the encoding rate in error-correction computation scenarios still requires further research. The trade-offs among different error-correction code schemes (surface code, toric code, LDPC codes, Bacon--Shor codes, etc.) in encoding rate, threshold, and overhead still need to be further verified experimentally, especially under the reconfigurable connectivity and parallel operation conditions characteristic of the neutral-atom platform. From a computer science perspective, different error-correction code schemes differ significantly in encoding rate, decoding complexity, and space--time overhead. The encoding rate of the surface code is about $1/(2d^2)$ (where $d$ is the code distance, i.e., the minimum-distance parameter corresponding to the number of errors the code can correct), decoding uses the minimum-weight perfect matching (MWPM) algorithm with complexity $O(d^2)$, but requires a large number of physical qubits (about $2d^2$ physical qubits for code distance $d$). The toric code has a similar encoding rate and decoding algorithm, but its threshold is slightly better under symmetric error models. The encoding rate of LDPC codes can approach 1/2 (such as the 2:1 memory encoding rate reported by QuEra), decoding uses belief propagation algorithms with complexity related to the sparsity of the parity-check matrix, and the space overhead is significantly lower than that of surface codes, but the connectivity requirements are higher---exactly satisfied by the all-to-all connectivity of the neutral-atom platform. The encoding rate of Bacon--Shor codes lies between the two; its subsystem code structure allows fewer measurements to detect errors, making it suitable for platforms with limited mid-circuit measurement capabilities.

In terms of specific mapping on the neutral-atom platform, the nearest-neighbor connections of the surface code can be directly mapped to a two-dimensional lattice arrangement of optical tweezers, while the long-range connections of LDPC codes require the use of atom movability to route and schedule remote entangling gates, placing higher algorithmic complexity demands on compiler qubit placement and gate scheduling optimization.

\subsection{Atom Loss and Mid-Circuit Replenishment}

The finite trap lifetime of atoms is an inherent limitation of neutral-atom quantum computing. Even though a 6000-second trap lifetime has been achieved in cryogenic environments, atom loss continues to occur in large-scale arrays. For large-scale quantum algorithms requiring execution times of several seconds to tens of seconds, atom loss will cause computational errors. Although atom replenishment technology is conceptually feasible, its engineering implementation faces challenges: the replenishment process needs to complete atom transfer and state initialization without disturbing qubits currently undergoing computation, and the replenished atoms need to precisely match the trapping conditions and quantum states of the original atoms. In 2025, Chiu et al. reported the continuous operation experiment of a 3000-qubit system, demonstrating for the first time on an engineering scale large-scale continuous atom replenishment---using a dual optical-lattice conveyor belt to deliver fresh atoms to the computing zone at a rate of 300,000 atoms per second while maintaining the coherence of already stored qubits, achieving continuous operation for more than two hours. In 2026, Bluvstein et al. proposed a loss-aware decoding scheme to address atom loss at the algorithmic level by incorporating it into the error-correction framework rather than completely eliminating it. The 17 $\mu$K sub-Doppler cooling and 17 cm light transmission technologies implemented by Infleqtion provide a hardware foundation for the continuous-operation architecture---in which the atom preparation region is spatially separated from the high-coherence computing region and fresh atoms are continuously delivered to the computing region. The gate-based readout and cooling scheme based on auxiliary qubits also demonstrates the use of Rydberg gates to realize conversion of quantum states from electronic degrees of freedom to motional degrees of freedom during computation, applicable to in-situ cooling. Achieving efficient real-time atom replenishment remains one of the core engineering challenges for large-scale fault-tolerant quantum computing.

\subsection{Laser Systems and Optical Engineering Industrialization}

Neutral-atom quantum computing places extremely high demands on laser systems: Rydberg excitation requires narrow-linewidth, high-power, high-stability ultraviolet lasers; optical tweezer arrays require precisely controlled laser phase and intensity distributions; and atom cooling and state preparation require coordinated multi-wavelength lasers. Currently, these laser systems are mostly customized in laboratories, with large size, complex debugging, and high environmental sensitivity. Integrating laser systems into industrial-grade products requires solving problems of laser miniaturization, optical-path integration, automatic alignment, and long-term stability. Integrated photonics technologies (such as silicon nitride optical waveguides and silicon photonic chips) may enable compact laser systems, but the technology of integrating them with atom trapping systems is still immature.

\subsection{Control Electronics and Scalability}

Parallel control of thousands or even tens of thousands of qubits poses severe challenges to electronic systems. The position and intensity of each optical tweezer need to be independently controlled, the Rydberg excitation of each atom requires precise pulse timing, and mid-circuit measurement requires high-speed feedback control. Current field-programmable gate array (FPGA) control schemes are feasible at the hundred-to-thousand-qubit scale, but expanding to above ten thousand qubits faces bottlenecks in channel count, bandwidth, and synchronization precision. In 2026, Wei et al. reported a 10 MHz refresh-rate spatial light modulator, increasing the control bandwidth of optical tweezer arrays by two orders of magnitude and serving as a core device for high-speed dynamic rearrangement of large-scale arrays. The MHz-rate two-dimensional optical potential scanning device developed by Bytyqi et al. (Vuletić group) provides an alternative scheme for high-speed atom manipulation. These technology breakthroughs are expected to alleviate the bandwidth bottleneck of control electronics, but developing application-specific integrated circuits (ASICs) or using time-multiplexing techniques to achieve fully parallel control above ten thousand qubits still requires substantial engineering investment.

From the perspective of computer system architecture, the unique ``move--entangle--separate'' operation mode of the neutral-atom platform places unique requirements on the instruction set architecture (ISA) and compiler. Unlike superconducting platforms with fixed coupling topologies, the connection relationships of the neutral-atom platform are dynamically generated---the compiler needs to decide in each circuit layer which qubit pairs should be brought close to execute entangling gates and which should be separated to avoid crosstalk. This ``qubit placement and routing'' problem is analogous to classical VLSI layout and routing, where the physical positions of qubits correspond to the placement of chip standard cells, the execution of entangling gates corresponds to the connectivity constraints of the routing network, and the blockade radius corresponds to the minimum spacing constraint in design rules. However, unlike fixed metal-layer routing in VLSI, the connection relationships of the neutral-atom platform are dynamically generated, and the computational complexity is higher in three-dimensional reconfigurable space. In addition, the neutral-atom platform supports parallel execution of multiple two-qubit gates in different regions of the array at the same time (as long as the gate pairs do not exceed the blockade radius of each other), which gives the compiler greater freedom for gate scheduling and parallelization optimization than fixed-topology platforms, but also increases the search space of scheduling algorithms. Current quantum compilation frameworks (such as Qiskit and Cirq) are mainly designed for fixed-topology platforms and lack native support for dynamic reconfigurable connections. Developing dedicated compilers and intermediate representations for the neutral-atom platform is an important direction for software stack construction. Recently, Huang et al. proposed the ZAP (Zoned Architecture and Performant compiler) for field-programmable atom arrays, dividing atom arrays into computing, storage, and readout zones to achieve collaborative optimization of atom routing and gate scheduling, providing a reference scheme for compiler design of large-scale neutral-atom platforms. At the programming model level, the neutral-atom platform supports both gate-model quantum computing and analog quantum computing modes; how to design unified programming abstractions to bridge the two modes and achieve efficient compilation of hybrid algorithms is also an urgent challenge at the software level.

\subsection{Long-Distance Quantum Interconnection}

Achieving distributed quantum computing and quantum networks requires connecting multiple neutral-atom quantum processors through photonic interconnection. The technology of the light--matter interface for neutral atoms---mapping atomic quantum states to photonic states and transmitting them between processors---is still in its early stages. Challenges include single-photon generation efficiency, photon--atom coupling efficiency, fiber transmission loss, and remote entanglement generation rate. Covey et al. systematically reviewed quantum network architectures with neutral-atom processing nodes, pointing out that coupling atom arrays with optical cavities can achieve efficient remote entanglement generation and is a feasible path for building distributed quantum computing systems. However, although individual experiments have demonstrated atom--photon entanglement and remote atomic entanglement, key indicators such as single-photon generation efficiency, photon--atom coupling efficiency, and remote entanglement generation rate still have large gaps from practical quantum interconnection. This bottleneck is not only a challenge faced by the neutral-atom platform but also a common problem for the entire quantum computing field.

\section{Outlook and Conclusion}

Neutral-atom quantum computing has experienced leapfrog development in the past five years, from hundred-atom-scale quantum simulators to ten-thousand-atom-scale arrays and thousand-qubit-scale logical quantum processors, and has become one of the most promising routes in quantum computing hardware. Based on the preceding analysis, the following conclusions can be drawn.

First, \textbf{optical tweezer arrays combined with Rydberg interactions} have become the mainstream technical route. Major domestic and international teams have adopted this scheme and achieved the most prominent results; its comprehensive advantages in scalability, gate fidelity, and reconfigurability have been fully verified. The optical lattice scheme still has unique value in quantum simulation and hybrid computing.

Second, \textbf{quantum error correction is moving from proof of principle to engineering implementation}. From logical quantum processors to algorithmic fault-tolerance frameworks and LDPC code schemes, and further to toric code multi-round error correction and 2:1 memory encoding rate breakthroughs, the path to fault-tolerant quantum computing is rapidly becoming concrete. The all-to-all connectivity and parallel operation capabilities of the neutral-atom platform are conducive to efficient quantum error correction.

Third, \textbf{scaling and engineering bottlenecks still exist}. Scaling technology from thousand-atom to ten-thousand-atom arrays is maturing, but bottlenecks such as breaking through gate fidelity above 99.9\%, bridging error-correction memory to error-correction computation, atom loss management, and laser system industrialization still require continuous research effort. By difficulty, these bottlenecks can be divided into three layers: atom loss management and array control bandwidth are near-term engineering bottlenecks with clear technical paths; gate fidelity limits and error-correction computation engineering are medium-to-long-term physical bottlenecks requiring breakthroughs in principles; and laser system integration and long-distance quantum interconnection may constitute hard constraints.

Fourth, \textbf{the domestic--international gap is narrowing but still exists}. China has reached an internationally leading level in key indicators such as atomic array scale and gate fidelity, but still lags behind the United States in experimental verification of quantum error correction, degree of industrialization, and ecosystem construction.

Neutral-atom quantum computing is expected to realize ten-thousand-qubit-scale physical quantum bit processors, hundred-qubit-scale logical qubit fault-tolerant computing, and practical demonstrations of quantum error correction advantage within the next 5--10 years. For China, it is recommended to strengthen layout in three aspects: quantum error correction experimental research, industry--academia--research collaboration, and interdisciplinary talent training, and to promote independent control of core devices.

\section*{Acknowledgements}
This work is supported by the National Key Research and Development Program of China under Grant No. 2023YFB4502500.

\section*{Innovation Statement}
The research problem addressed in this paper belongs to the neutral-atom quantum computing technology route review in the field of quantum computing hardware. Neutral-atom quantum computing uses laser-trapped neutral atoms as qubits and implements quantum logic gate operations through Rydberg-state interactions. Because of its excellent scaling potential, it is regarded as one of the most promising technology routes for practical fault-tolerant quantum computing.

Internationally, this field has experienced explosive development from 2021 to 2026: atomic array scale has expanded from tens to the ten-thousand level (stable trapping of about 11,000 atoms was achieved in 2026), two-qubit gate fidelity has increased to 99.84\%, and quantum error correction has advanced from proof of principle to multi-round continuous error correction based on the toric code and a 2:1 memory encoding rate breakthrough. This paper systematically reviews the full-chain technological progress from the theoretical foundation of Rydberg blockade in 2000 to the ten-thousand-atom array in 2026, and analyzes it in four dimensions: ``physical basis---hardware scale---error correction and fault tolerance---application exploration.'' It also provides a comparative analysis of the domestic and international industrialization landscape.

Compared with existing reviews, the features and innovations of this paper are: (1) it covers the latest breakthrough achievements from 2025 to 2026, including metasurface optical tweezer arrays, algorithmic fault-tolerance frameworks, and loss-aware decoding, with a more up-to-date time span; (2) it systematically analyzes the leap from quantum error correction memory to computation, clearly pointing out the key limitation that the current 2:1 encoding rate applies only to memory functions; (3) it quantifies the physical limits of gate fidelity improvement through theoretical analysis and provides risk commentary on domestic and international industrial roadmaps; (4) it systematically reviews progress in domestic and international academia and industry.


\begin{thebibliography}{99}

\bibitem{Bluvstein2024} Bluvstein D, Levine H, Semeghini G, et al. Logical quantum processor based on reconfigurable atom arrays[J]. \textit{Nature}, 2024, 626: 58--65

\bibitem{Lin2025} Lin R, Zhong H S, Li Y, et al. AI-enabled parallel assembly of thousands of defect-free neutral atom arrays[J]. \textit{Physical Review Letters}, 2025, 135: 063401

\bibitem{Evered2023} Evered S J, Bluvstein D, Kalinowski M, et al. High-fidelity parallel entangling gates on a neutral-atom quantum computer[J]. \textit{Nature}, 2023, 622: 268--272

\bibitem{Zhou2025} Zhou H, Zhao C, Li S H, et al. Low-overhead transversal fault tolerance for universal quantum computation[J]. \textit{Nature}, 2025, 646: 303--308

\bibitem{Microsoft2024} Microsoft Azure Quantum, Atom Computing. Microsoft and Atom Computing offer a commercial quantum machine with the largest number of entangled logical qubits on record[EB/OL]. (2024-11-19)[2026-06-19]. \url{https://azure.microsoft.com/en-us/blog/quantum/2024/11/19/microsoft-and-atom-computing-offer-a-commercial-quantum-machine-with-the-largest-number-of-entangled-logical-qubits-on-record}

\bibitem{Pasqal2024} Pasqal. Pasqal exceeds 1,000 atoms in quantum processor[EB/OL]. (2024-06)[2026-06-19]. \url{https://www.pasqal.com}

\bibitem{Pecorari2025} Pecorari L, Witzel W, Goldman M, et al. High-rate quantum LDPC codes for long-range-connected neutral atom registers[J]. \textit{Nature Communications}, 2025, 16: 1111

\bibitem{Ruiz2025} Ruiz D, Guillaud J, Leverrier A, et al. LDPC-cat codes for low-overhead quantum computing in 2D[J]. \textit{Nature Communications}, 2025, 16: 1040

\bibitem{Bravyi2024} Bravyi S, Cross A W, Yoder T J, et al. High-threshold and low-overhead fault-tolerant quantum memory[J]. \textit{Nature}, 2024, 627: 778--782

\bibitem{Jaksch2000} Jaksch D, Cirac J I, Zoller P, et al. Fast quantum gates for neutral atoms[J]. \textit{Physical Review Letters}, 2000, 85: 2208

\bibitem{Saffman2005} Saffman M, Walker T G. Analysis of a quantum logic device based on dipole-dipole interactions of optically trapped Rydberg atoms[J]. \textit{Physical Review A}, 2005, 72: 022347

\bibitem{Saffman2010} Saffman M, Walker T G, M{\o}lmer K. Quantum information with Rydberg atoms[J]. \textit{Reviews of Modern Physics}, 2010, 82: 2313--2363

\bibitem{Urban2009} Urban E, Johnson T A, Henage T, et al. Observation of Rydberg blockade between two atoms[J]. \textit{Nature Physics}, 2009, 5: 110--114

\bibitem{Gaetan2009} Ga{\"e}tan A, Miroshnychenko Y, Wilk T, et al. Observation of collective excitation of two individual atoms in the Rydberg blockade regime[J]. \textit{Nature Physics}, 2009, 5: 115--118

\bibitem{Isenhower2010} Isenhower L, Urban E, Zhang X L, et al. Demonstration of a neutral atom controlled-NOT quantum gate[J]. \textit{Physical Review Letters}, 2010, 104: 010503

\bibitem{Labuhn2016} Labuhn H, Barredo D, Ravets S, et al. Tunable two-dimensional arrays of single Rydberg atoms for realizing quantum Ising models[J]. \textit{Nature}, 2016, 534: 667--670

\bibitem{Barredo2018} Barredo D, Lienhard V, de Leseleuc S, et al. Synthetic three-dimensional atomic structures assembled atom by atom[J]. \textit{Nature}, 2018, 561: 79--82

\bibitem{Bernien2017} Bernien H, Schwartz S, Keesling A, et al. Probing many-body dynamics on a 51-atom quantum simulator[J]. \textit{Nature}, 2017, 551: 579--584

\bibitem{Ebadi2021} Ebadi S, Wang T, Levine H, et al. Quantum phases of matter on a 256-atom programmable quantum simulator[J]. \textit{Nature}, 2021, 595: 227--232

\bibitem{Scholl2021} Scholl P, Schuler M, Williams H J, et al. Quantum simulation of 2D antiferromagnets with hundreds of Rydberg atoms[J]. \textit{Nature}, 2021, 595: 233--238

\bibitem{Semeghini2021} Semeghini G, Levine H, Keesling A, et al. Probing topological spin liquid on a programmable quantum simulator[J]. \textit{Science}, 2021, 374: 1242--1247

\bibitem{Levine2018} Levine H, Keesling A, Semeghini G, et al. High-fidelity control and entanglement of Rydberg-atom qubits[J]. \textit{Physical Review Letters}, 2018, 121: 123603

\bibitem{Graham2022} Graham T M, Song Y, Scott J, et al. Multi-qubit entanglement and algorithms on a neutral-atom quantum computer[J]. \textit{Nature}, 2022, 604: 457--462

\bibitem{Schymik2021} Schymik K N, Pancaldi S, Nogrette F, et al. Single atoms with 6,000-second trapping lifetimes in optical-tweezer arrays at cryogenic temperatures[J]. \textit{Physical Review Applied}, 2021, 16: 034013

\bibitem{Xu2024} Xu Q, Ataides-Patino J P, Brown H, et al. Constant-overhead fault-tolerant quantum computation with reconfigurable atom arrays[J]. \textit{Nature Physics}, 2024, 20: 1084--1090

\bibitem{Rodriguez2025} Rodriguez P S, Robinson J M, Jepsen P N, et al. Experimental demonstration of logical magic state distillation[J]. \textit{Nature}, 2025, 645: 620--625

\bibitem{Bluvstein2026} Bluvstein D, Evered S J, Geim A A, et al. A fault-tolerant neutral-atom architecture for universal quantum computation[J]. \textit{Nature}, 2026, 649: 39--46

\bibitem{Zhao2026} Zhao C, Duckering C, Gu A, et al. Towards ultra-high-rate quantum error correction with reconfigurable atom arrays[J]. arXiv preprint, arXiv:2604.16209, 2026

\bibitem{Okada2026} Okada K, Kasai K. High-girth regular quantum LDPC codes from affine-coset structures[J]. arXiv preprint, arXiv:2604.20838, 2026

\bibitem{Atom2026} Atom Computing. Atom Computing reveals quantum error correction with toric code[EB/OL]. (2026-06-03)[2026-06-19]. \url{https://www.atom-computing.com}

\bibitem{Wei2026} Wei X, Li Z, Karve A V, et al. A 10 megahertz spatial light modulator[J]. arXiv preprint, arXiv:2601.08906, 2026

\bibitem{Bytyqi2026} Bytyqi E, Sinclair J, Ramette J, et al. Device for MHz-rate rastering of arbitrary 2D optical potentials[J]. arXiv preprint, arXiv:2602.16025, 2026

\bibitem{Tsai2026} Tsai R B-S, Picard L R B, Sun X, et al. Gate-based readout and cooling of neutral atoms[J]. arXiv preprint, arXiv:2603.21643, 2026

\bibitem{Stein2026} Stein S, Liu C, Kan S, et al. Multitarget Rydberg gates via spatial blockade engineering[J]. \textit{Physical Review Research}, 2026, 8: 013254

\bibitem{Li2026} Li R, Zhang M H, Qian J. Optimized ancillary drive for fast Rydberg entangling gates[J]. \textit{Physical Review A}, 2026, 113: 032614

\bibitem{White2026} White R, Ramesh V, Impertro A, et al. Quantum cellular automata on a dual-species Rydberg processor[J]. arXiv preprint, arXiv:2601.16257, 2026

\bibitem{Infleqtion2026} Infleqtion. Infleqtion strengthens neutral-atom quantum computing platform with new technical breakthroughs[EB/OL]. (2026-05-20)[2026-06-19]. \url{https://www.infleqtion.com}

\bibitem{Lin2026} Lin R, Li Y, Zheng L T, et al. Sustaining high-fidelity quantum logic in neutral-atom circuits via mid-circuit operations[J]. arXiv preprint, arXiv:2603.01612, 2026

\bibitem{PhotonBox2026} Photon Box Research Institute. 2026 global quantum computing industry development outlook[EB/OL]. (2026-02)[2026-06-19]. \url{https://www.photonboxonline.com}

\bibitem{Manetsch2025} Manetsch H J, Nomura G, Bataille E, et al. A tweezer array with 6100 highly coherent atomic qubits[J]. \textit{Nature}, 2025, 647: 60--67

\bibitem{Chiu2025} Chiu N C, Trapp E C, Guo J, et al. Continuous operation of a coherent 3,000-qubit system[J]. \textit{Nature}, 2025, 646: 1075--1080

\bibitem{Evered2025} Evered S J, Kalinowski M, Geim A A, et al. Probing the Kitaev honeycomb model on a neutral-atom quantum computer[J]. \textit{Nature}, 2025, 645: 341--347

\bibitem{Wang2023} Wang S, Zhang W, Zhang T, et al. Accelerating the assembly of defect-free atomic arrays with maximum parallelisms[J]. \textit{Physical Review Applied}, 2023, 19: 054032

\bibitem{Zhang2026} Zhang Z, Wang Y, Liao Y, et al. Direct generation of an array with 78,400 optical tweezers using a single metasurface[J]. \textit{Chinese Physics Letters}, 2026, 43: 010606

\bibitem{Cheng2024} Cheng Y, Zhai H. Emergent U(1) lattice gauge theory in Rydberg atom arrays[J]. \textit{Nature Reviews Physics}, 2024, 6: 566--576

\bibitem{Yang2016} Yang J, He X, Guo R, et al. Coherence preservation of a single neutral atom qubit transferred between magic-intensity optical traps[J]. \textit{Physical Review Letters}, 2016, 117: 123201

\bibitem{Zeng2017} Zeng Y, Xu P, He X, et al. Entangling two individual atoms of different isotopes via Rydberg blockade[J]. \textit{Physical Review Letters}, 2017, 119: 160502

\bibitem{Qiu2020} Qiu X, Zoller P, Li X. Programmable quantum annealing architectures with Ising quantum wires[J]. \textit{PRX Quantum}, 2020, 1: 020311

\bibitem{Ye2023} Ye M, Tian Y, Lin J, et al. Universal quantum optimization with cold atoms in an optical cavity[J]. \textit{Physical Review Letters}, 2023, 131: 103601

\bibitem{Cao2025} Cao S, Li X. Hardware-efficient Rydberg atomic quantum solvers for NP problems[J/OL]. arXiv preprint, arXiv:2507.22686, 2025

\bibitem{Li2025} Li X, Hou J Y, Wang J C, et al. A fiber array architecture for atom quantum computing[J]. \textit{Nature Communications}, 2025, 16: 9728

\bibitem{Beaulieu2024} Beaulieu D, Kornja{\v c}a M, Krunic Z, et al. Robust quantum reservoir computing for molecular property prediction[J]. arXiv preprint, arXiv:2412.06758, 2024

\bibitem{Matwiejew2025} Matwiejew E, Wurtz J, Chen J, et al. Continuous-time quantum walk-based ans{\"a}tze on neutral atom hardware[J]. arXiv preprint, arXiv:2509.00386, 2025

\bibitem{Ismail2026} Ismail R, Chen I, Zhao C, et al. Transversal architecture for megaquop-scale quantum simulation with neutral atoms[J]. \textit{PRX Quantum}, 2026, 7: 020343

\bibitem{Norrell2026} Norrell S A, Shen Y, Saffman M, et al. Entangling gate performance and fidelity limits with neutral atom F{\"o}rster resonances[J]. arXiv preprint, arXiv:2605.19245, 2026

\bibitem{QuEra2024} QuEra Computing. QuEra Computing releases a groundbreaking roadmap for advanced error-corrected quantum computers[EB/OL]. (2024-01-09)[2026-06-19]. \url{https://www.quera.com/press-releases/quera-computing-releases-a-groundbreaking-roadmap-for-advanced-error-corrected-quantum-computers-pioneering-the-next-frontier-in-quantum-innovation-0}

\bibitem{Atom2025} Atom Computing. Atom Computing and Microsoft accelerate commercial quantum computing with next-generation fault-tolerant quantum computer[EB/OL]. (2025-01-20)[2026-06-19]. \url{https://atom-computing.com/microsoft-partnership}

\bibitem{Pasqal2025} Pasqal. Pasqal releases 2025 roadmap showcasing upgradable platform from today's quantum solutions to tomorrow's fault-tolerant systems[EB/OL]. (2025-06-12)[2026-06-19]. \url{https://www.pasqal.com/newsroom/pasqal-releases-2025-roadmap}

\bibitem{Sun2020} Sun Y, Xu P, Chen P X, Liu L. Controlled phase gate protocol for neutral atoms via off-resonant modulated driving[J]. \textit{Physical Review Applied}, 2020, 13: 024059

\bibitem{Fu2022} Fu Z, Xu P, Sun Y, et al. High-fidelity entanglement of neutral atoms via a Rydberg-mediated single-modulated-pulse controlled-PHASE gate[J]. \textit{Physical Review A}, 2022, 105: 042430

\bibitem{Sun2023} Sun Y. Suppression of high-frequency components in off-resonant modulated driving protocols for Rydberg-blockade gates[J]. \textit{Physical Review Applied}, 2023, 20: L061002

\bibitem{Rines2025} Rines R, Hall B, First M H, et al. Demonstration of a logical architecture uniting motion and in-place entanglement: Shor's algorithm, constant-depth CNOT ladder, and many-hypercube code[J]. arXiv preprint, arXiv:2509.13247, 2025

\bibitem{Wang2026} Wang Y, Zhang Z, Zhang T, et al. Trapping 11,000 atoms in a tweezer array generated by a single metasurface[J]. arXiv preprint, arXiv:2606.02715, 2026

\bibitem{Fowler2012} Fowler A G, Mariantoni M, Martinis J M, et al. Surface codes: Towards practical large-scale quantum computation[J]. \textit{Physical Review A}, 2012, 86: 032324

\bibitem{Barnes2022} Barnes K, Battaglino P, Bloom B, et al. Assembly and coherent control of a register of nuclear spin qubits[J]. \textit{Nature Communications}, 2022, 13: 2779

\bibitem{Deist2022} Deist E, Lu Y A, Ho J, et al. Mid-circuit cavity measurement in a neutral atom array[J]. \textit{Physical Review Letters}, 2022, 129: 203602

\bibitem{Graham2023} Graham T M, Phuttitarn M, Chinnarasu R, et al. Midcircuit measurements on a single-species neutral alkali atom quantum processor[J]. \textit{Physical Review X}, 2023, 13: 041051

\bibitem{Fudan2025} Fudan University. Large-scale atomic quantum computing system jointly developed by young Fudan scientists[EB/OL]. (2025-08-30)[2026-06-25]. \url{https://news.fudan.edu.cn/2025/0830/c1247a146502}

\bibitem{Zhongke2026} Zhongke Kuyuan Technology (Wuhan) Co., Ltd. Hanyuan series neutral atom quantum computer[EB/OL]. [2026-06-25]. \url{https://www.quantumchina.com/newsinfo/7282381.html}

\bibitem{Taiyi2026} Taiyi Liangsheng (Shanghai) Quantum Technology Co., Ltd. Ytterbium atom neutral atom quantum computing platform[EB/OL]. (2026-03-17)[2026-06-25]. \url{https://finance.eastmoney.com/a/202603173673595280}

\bibitem{Awschalom2018} Awschalom D D, Hanson R, Wrachtrup J, et al. Quantum technologies with optically interfaced solid-state spins[J]. \textit{Nature Photonics}, 2018, 12(9): 516--527

\bibitem{Huang2026} Huang C, Zhao X, Xu H, et al. ZAP: Zoned architecture and performant compiler for field programmable atom array[J]. \textit{IEEE Transactions on Quantum Engineering}, 2026. arXiv preprint, arXiv:2411.14037

\bibitem{Covey2023} Covey J P, Weinfurter H, Bernien H. Quantum networks with neutral atom processing nodes[J]. \textit{npj Quantum Information}, 2023, 9: 90

\bibitem{Maller2015} Maller K M, Lichtman M T, Xia T, et al. Rydberg-blockade controlled-NOT gate and entanglement in a two-dimensional array of neutral-atom qubits[J]. \textit{Physical Review A}, 2015, 92: 022336

\bibitem{Saffman2025} Saffman M. Quantum computing with atomic qubit arrays: confronting the cost of connectivity[J]. arXiv preprint, arXiv:2505.11218, 2025

\end{thebibliography}
\end{document}